\documentclass{ieeetj}
\usepackage{cite}
\usepackage{amsmath,amssymb,amsfonts}
\usepackage{algorithmic}
\usepackage{graphicx,color}
\usepackage{textcomp}
\usepackage{xcolor}
\usepackage{hyperref}
\usepackage{enumerate}
\usepackage{multirow}
\usepackage{gensymb}
\usepackage{booktabs}
\usepackage{amsmath}
\usepackage{siunitx}
\usepackage[final]{changes}
\hypersetup{hidelinks=true}
\usepackage{algorithm,algorithmic}

\def\BibTeX{{\rm B\kern-.05em{\sc i\kern-.025em b}\kern-.08em
    T\kern-.1667em\lower.7ex\hbox{E}\kern-.125emX}}
\AtBeginDocument{\definecolor{tmlcncolor}{cmyk}{0.93,0.59,0.15,0.02}\definecolor{NavyBlue}{RGB}{0,86,125}}

\def\OJlogo{\vspace{-4pt}$<$Society logo(s) and publication title will appear here.$>$}
\def\seclogo{\vspace{10pt}$<$Society logo(s) and publication title will appear here.$>$}
\def\OJlogo{\vspace{-4pt}}
\def\seclogo{\vspace{10pt}}

\def\authorrefmark#1{\ensuremath{^{\textbf{#1}}}}
\usepackage[nolist]{acronym}
\begin{document}
\newacro{DOA}[DOA]{Direction of Arrival}
\newacro{AVSL}[AVSL]{Audio-visual Speaker Localization}
\newacro{TDOA}[TDOA]{Time Difference of Arrival}
\newacro{CE}[CE]{Cross Entropy}
\newacro{FOV}[FOV]{Field of View}
\newacro{KF}[KF]{Kalman Filter}

\markboth{Audio-Visual Speaker Tracking: Progress, Challenges, and Future Directions}{ZHAO {ET AL.} }

\title{Audio-Visual Speaker Tracking: Progress, Challenges, and Future Directions}

\author{JINZHENG ZHAO\authorrefmark{1}, YONG XU\authorrefmark{2}, (Senior Member, IEEE), \\XINYUAN QIAN\authorrefmark{3}, (Senior Member, IEEE), DAVIDE BERGHI\authorrefmark{1}, (Student Member, IEEE),\\PEIPEI WU\authorrefmark{1}, MENG CUI\authorrefmark{1}, JIANYUAN SUN\authorrefmark{4}, PHILIP J.B. JACKSON\authorrefmark{1}, (Member, IEEE), WENWU WANG\authorrefmark{1}, (Senior Member, IEEE)}
\affil{Centre for Vision, Speech and Signal Processing, University of Surrey, Guildford, GU2 7XH, UK}
\affil{Tencent AI Lab, Bellevue, WA 98004, USA}
\affil{Department of Computer Science and Technology, University of Science and Technology Beijing, Beijing 100083, China}
\affil{Department of Computer Science, University of Sheffield, Sheffield, S1 4DP,
UK}
\corresp{Corresponding author: Jinzheng Zhao (email: j.zhao@surrey.ac.uk).}
\authornote{This research is supported by Tencent AI Lab Rhino-Bird Gift Fund and University of Surrey.}

\begin{abstract}
Audio-visual speaker tracking has drawn increasing attention over the past few years due to its academic values and wide applications. Audio and visual modalities can provide complementary information for localization and tracking. With audio and visual information, the Bayesian-based filter and deep learning-based methods can solve the problem of data association, audio-visual fusion and track management. 
 In this paper, we conduct a comprehensive overview of audio-visual speaker tracking. To our knowledge, this is the first extensive survey over the past five years. We introduce the family of Bayesian filters and summarize the methods for obtaining audio-visual measurements. In addition, the existing trackers and their performance on the AV16.3 dataset are summarized. In the past few years, deep learning techniques have thrived, which also boost the development of audio-visual speaker tracking. The influence of deep learning techniques in terms of measurement extraction and state estimation is also discussed. Finally, we discuss the connections between audio-visual speaker tracking and other areas such as speech separation and distributed speaker tracking.
\end{abstract}



\maketitle

\section{INTRODUCTION}
\label{intro}
The goal of audio-visual speaker tracking is to determine the positions of the speaker in each time step using data from sensors like microphones and cameras. It has wide applications, including but not limited to human-computer interaction \cite{shivappa2010audio}, speech recognition \cite{potamianos2003joint}, speaker diarization \cite{gebru2017audio}, speech enhancement \cite{loizou2007speech} and surveillance \cite{hampapur2005smart}. In addition, it has been used to automatically extract tracking metadata for object-based media production \cite{pike2016object-based,Coleman:2018:objectBased}, where audio-visual objects are faithfully spatialized according to their position in space \cite{izhar:2020:AVtracker,Schweiger:2022:tool6dof}. 

With audio, speaker location can be obtained in omnidirection (\added{except in linear or planar arrays due to front-back ambiguity \cite{hohnerlein2017perceptual} or with directional microphones}), albeit in relatively low resolution. In comparison, the localization resolution offered by visual signals is often higher, but the localization can only be achieved when the speaker is in the field of view of the cameras. Thus, audio provides a complementary modality to overcome limitations of visual modality under conditions such as occlusion, field-of-view constraints, and poor illumination where visual cues degrade. In contrast, when audio is affected by strong room reverberation and ambient noise, video information can serve as a backup. This indicates the collaborative potential of multiple modalities to improve tracking performance. 

The audio-visual multi-speaker tracking task presents several challenges that require careful consideration, including (1) integrating audio and visual data in a complementary manner, (2) estimating the number of simultaneous speakers which is unknown and dynamically changing \cite{zhao2022audio}, (3) dealing with many complex sources of uncertainty, such as missing detections, noise, clutter, and absent modality, and (4) improving tracking efficiency while maintaining tracking accuracy.

Various methods have been developed to address these challenges, as discussed in a survey paper \cite{kilicc2017audio}. However, there is a lack of review of emerging methods published in recent five years. The aim of this survey paper is to provide a comprehensive review of audio-visual speaker tracking, with a focus on emerging methods in recent five years. We provide a comprehensive and up-to-date literature review including visual measurements, audio measurements generation, Bayesian trackers, datasets and metrics. The methods discussed in \cite{kilicc2017audio} are mostly statistics-based methods, while we include more learning-based techniques in our survey, such as learning-based audio-visual measurements and learning-based tracking methods including deep learning and differentiable particle filters.

\added{The content covered in this survey includes five aspects, as presented in the following sections. First, we discuss visual measurements and audio measurements often used in audio-visual speaker tracking systems, including parametric-based methods and learning-based methods. Secondly, we present the tracking algorithms, including deep-learning based trackers, Bayesian filters, and differentiable Bayesian filters. Thirdly, we analyze important modules in the tracking system, such as multi-modal fusion and data association. Fourthly, we present the commonly-used dataset and evaluation metrics. Fifthly, we compare the performance of difference trackers, and analyze their advantages, limitations, and relations. }

\section{Measurements}
\label{measurements}
\begin{figure}[tbp]
    \centering
    \includegraphics[width=0.85\columnwidth]{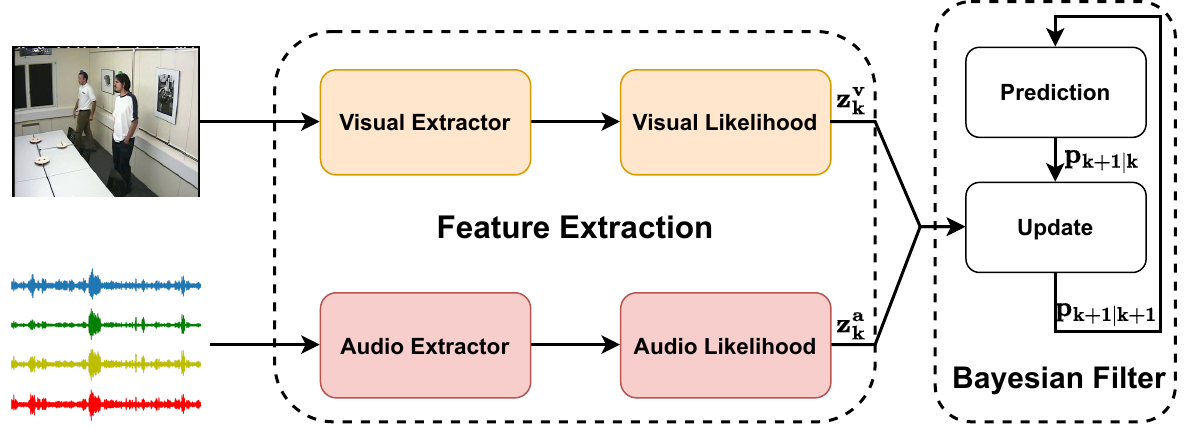}
    \caption{The structure of audio-visual speaker tracker.}
    \label{flow}
\end{figure}

\added{Measurements are used to correct the estimation in the transition step by the Bayesian filter.}
Both audio and visual modalities can serve as measurements for tracking, as depicted in Figure \ref{flow}. As pointed out by \cite{qian2021audio}, the measurement likelihoods can be classified as generative or discriminative. The former calculates the possibility map in the feature space and finds the most possible regions where speakers will appear, which can be regarded as a similarity matching problem. The latter often employs a pretrained detector to locate speakers and give coordinates as direct measurements.

\subsection{Visual Measurements}

Visual modality is superior to audio modality in terms of localization accuracy as it can offer richer information. Face detectors and color histograms are often employed to extract features from images.

\subsubsection{Parametric-Based Methods}

Color histogram methods can provide generative visual likelihoods by comparing the similarities between the reference image and the whole image search space. The reference images are often selected from the initial frame where speakers appear in a consecutive sequence. Color spatiogram \cite{birchfield2005spatiograms} is an alternative \cite{qian20173d, qian20183d}, which is enhanced by spatial means and covariance for each histogram bin to give a richer representation. Color histogram methods have been widely used \cite{kilicc2014audio, kilicc2016mean} to provide visual measurements. In scenarios where face detectors fail, color histogram methods can provide similarity feature maps as complementary information. The commonly used color representation is RGB and HSV. The similarity between two HSV histograms is calculated based on the Bhattacharyya distance:
\begin{equation}
    D=\sqrt{1-\sum_{n=1}^{N} \sqrt{r(n) q(n)}}
    \label{eq2}
\end{equation}

\noindent where $N$ is the number of histogram bins, $r(n)$ is the Hue histogram of the reference image. The reference image is often selected as the initial frame where the speaker is visible. $q(n)$ is the Hue histogram of the search area.

\subsubsection{Learning-Based methods}

As deep learning technologies thrive, face detectors are becoming quicker, stronger, and more robust. They provide coordinates of speakers' faces and discriminative likelihoods. MXNet \cite{wu2018light} was used in \cite{qian2019multi} and \cite{qian2021audio} for detection and the dual shot face detector (DSFD) \cite{li2019dsfd} was employed in \cite{zhao2022audio} to provide mouth positions. 

In addition to face detectors, object detectors like SSD \cite{liu2016ssd} and person detectors like \cite{cao2017realtime} can also be used in audio-visual tracking \cite{wilson2020avot, ban2019variational}. After obtaining the coordinates of bounding boxes, some work \cite{qian2022audio, wang2022nerc} encodes them to Gaussian vectors which represent the posterior distribution of the object positions along the horizontal and vertical axis.


Another option is to provide a generative visual likelihood with a learning method, such as the Siamese Network \cite{bromley1993signature}. Similar to color histograms, the similarity between the reference image and the search area is calculated on the learned features extracted with deep learning methods. For example, \cite{li2021multi} adopts a pretrained fully-convolutional siamese network \cite{bertinetto2016fully} to calculate the response map which is then used as the visual measurement.

\subsubsection{Comparison between Parametric-Based Methods and Learning-Based Methods}
\added{Parametric-based methods are training-free and easy to implement. It is suitable for simple applications. However, the performance degrades in challenging scenarios such as bad lightning and different speaker orientations. In contrast, learning-based methods have high accuracy but suffer from high computational cost and sometimes low inference speed.}

\subsection{Audio Measurements}

Audio-visual speaker tracking relies on sound source localization (SSL) algorithms to obtain measurements by detecting and localizing the active sound sources. SSL methods can be broadly categorized into two groups: parametric-based methods and learning-based methods. 
\subsubsection{Parametric-Based Methods}
Parametric-based methods typically rely on Time Difference of Arrival (TDOA) estimation \cite{knapp1976generalized}, which often requires a computationally expensive global maximum search \cite{Do:2007:RT_SRP-PHAT}. They work well under general conditions but tend to fail in the presence of strong reverberation and noise.

\paragraph{Global Coherence Field}


Estimating the TDOA between different microphones provides useful spatial information for localization. Generalized Cross Correlation (GCC) is used for TDOA estimation but struggles when encountering background noise and room reverberation. To mitigate this problem, compared to GCC, Generalized Cross Correlation with Phase Transform (GCC-PHAT) is normalized by the magnitude while retaining the phase information, which is more robust under a bad environment \cite{omologo1997use}. Global coherence field (GCF) \cite{omologo1998spoken} gathers the spatial information by adding up GCC-PHAT of all microphone pairs. Peaks in the GCF map indicate the most likely position of the dominant speaker.

The computation of GCF involves two steps. Firstly, GCC-PHAT is calculated for the $j$-th pair of audio recorded in the microphone array, denoted as $s_{j} \in S$, at time $t$. This calculation is defined as follows:
\begin{equation}
  G_{j}(\tau, t)=\int_{-\infty}^{+\infty} \frac{\mathcal{F}_{s_{j,1}}(t, f) \mathcal{F}_{s_{j,2}}^{*}(t, f)}{\left|\mathcal{F}_{s_{j, 1}}(t, f)\right|\left|\mathcal{F}_{s_{j, 2}}^{*}(t, f)\right|} e^{j 2 \pi f \tau} d f
  \label{eq3}
\end{equation}

\noindent where $\tau$ represents the inter-microphone time lag, $f$ denotes the frequency,  $\mathcal{F}$ is short for the Short-Time Fourier Transform, $s_{j, 1}$ and $s_{j, 2}$ are the two microphones within the $j$-th pair, and $*$ signifies the complex conjugate. By summing the GCC-PHAT values across all pairs with the number of $|S|$, the final GCF is obtained.
\begin{equation}
  GCF(\boldsymbol{p}, t)=\frac{1}{|S|} \sum_{n=1}^{|S|} G_{n}\left(\tau_{n}(\boldsymbol{p}), t\right)
  \label{eq4}
\end{equation}
\noindent where $\boldsymbol{p}$ denotes discrete points sampled in the search space. \added{$\tau_{n}(\boldsymbol{p})$ is the TDOA for microphone pair $n$ if the sound source is in $\boldsymbol{p}$.} The discrete point resulting in the maximum of GCF is regarded as the sound source.




\paragraph{GCC-PHAT de-emphasis}
GCC-PHAT de-emphasis \cite{brutti2008localization} is proposed for adapting GCC-PHAT to scenarios of multiple speakers. After localizing the dominant speaker using GCF, the time lag corresponding to the dominant speaker is masked and GCF is re-calculated using the masked GCC-PHAT for localizing the non-dominant speakers. However, as indicated in \cite{qian2021audio}, GCC-PHAT de-emphasis does not perform well with the increasing number of speakers. And as shown in \cite{zhao2022audio2}, even in the two-speaker scenario, the performance of GCC-PHAT de-emphasis is not satisfactory when the speakers are close to each other.

\paragraph{stGCF}
There are additional GCF derivatives. For instance, \cite{li2021multi} proposed space-temporal GCF (stGCF), which inserts spatial and temporal information assisted by visual modality and improves localization accuracy. 

Assume $\mathbf{Q}^{2d} = \{q_{11}^{2d}, ..., q_{wh}^{2d}\}$ is the sampling points across the image plane. Through the camera projection model\cite{hartley2003multiple}, the 2D sampling points can be converted to groups of 3D points $\mathbf{Q}^{3d}_{k} = \{q_{11k}^{3d}, ..., q_{whk}^{3d}\}$ with different depths $\mathbf{D} = \{d_{1}, ...,d_{k}, ..., d_{L}\}$. 
\begin{equation}
\mathbf{Q}^{3d}_{k} = \Phi(\mathbf{Q}^{2d}, \added{d_{k}})
\end{equation}
Then we obtain the GCF in different depth $GCF(\mathbf{Q}^{3d}_{k}, t)$. The spatial GCF is defined as $GCF(\mathbf{Q}^{3d}_{k_{m}}, t)$ where the maximum of $GCF$ is achieved on the $k_{m}$-th depth.
The spatial GCF is obtained over frames in $[t -n_{1}]$. Then the first $n_{2}$ largest spatial GCF is selected as spatial-temporal GCF.

Apart from GCF, other algorithms such as MUSIC \cite{schmidt1986multiple}, independent component analysis \cite{sawada2005multiple}, and logistic regression \cite{trowitzsch2019joining} can also be used for providing audio measurements.

\subsubsection{Learning-Based Methods}
Learning-based methods are emerging as deep learning techniques thrive, which predict DOA \cite{adavanne2018direction} or Cartesian coordinates \cite{vera2018towards} \cite{Berghi:2021:mmsp} through neural networks trained to learn the mapping function that relates audio input features to the sound source positions. Properly trained learning-based methods tend to generalize well even when the signal-to-noise ratio (SNR) is low or in highly reverberant environments \cite{Xiao:2015:learning-based}.

With deep learning methods emerging, an increasing number of works tackle the problem of sound source localization by training an audio network, which aims to find the relationship between the input audio features and the positions of the sound sources. Compared to traditional parametric-based methods, the learning-based methods are more robust and generalize better in the presence of reverberation and acoustic noise. Neural networks were employed in \cite{vera2018towards} to predict the 3D positions of speakers given multi-channel audio signals recorded by microphone arrays. A DeepGCC was designed in \cite{vera2021acoustic} that can estimate sound source positions robustly in different environments with various room geometry and microphone array configurations. A convolutional recurrent neural network (CRNN) was used in \cite{adavanne2018sound} to detect and localize sound events concurrently. CRNN has been widely employed to localize moving sounds \cite{adavanne2019localization}, and with different audio input features \cite{adavanne2017sound,Cordourier2019GCCPHATCA, Berghi:2023:WASPAA}.
In \cite{cao2019polyphonic}, a two-stage strategy was used where sound event detection is conducted first, and then the predicted event is used to assist the localization of sound sources. In \cite{schymura2021exploiting}, a sequence-to-sequence model with an attention mechanism was designed to predict DOA. In \cite{schymura2021pilot}, Transformer \cite{vaswani2017attention} is used for localizing sound sources.

The audio features Mel-spectrogram, the mel frequency cepstral coefficients (MFCC), and GCC-PHAT are often selected as the input. Nguyen et al. \cite{nguyen2021salsa} recently proposed the Spatial Cue-Augmented Log-Spectrogram (SALSA) features. They consist of a normalized version of the principal eigenvector of the spatial covariance matrix computed at each time-frequency bin. This enables concatenation with the spectrograms extracted from the microphone array's channels.
Subsequently, the authors proposed SALSA-Lite \cite{nguyen2021salsaa}, a lighter version consisting of the frequency-normalized interchannel phase difference (IPD) computed at each time-frequency bin. These features have shown promising performance on task 3 of the Challenge on Detection and Classification of Acoustic Scenes and Events (DCASE): sound event detection and localization. 

\subsubsection{Comparison between Parametric-Based Methods and Learning-Based Methods}
\added{Most of parametric-based methods are based on TDOA estimation and beamforming, which is interpretable and explainable. In addition, it does not need training data and saves computational resources. However, it is sensitive to noise and reverberation. And the model performance is specific on the microphone array geometry.}

\added{The learning-based methods can adapt to different kinds of microphone arrays, noise and reverberation through training. But it requires large amounts of labeled data.}

\subsubsection{Extracting Features in Teacher-Student Paradigm}
\label{teacherstudent}
As summarized in the last section, learning-based sound source localization methods typically require extensive amounts of annotated training data. However, in the task of audio-visual speaker tracking, acquiring such data is often challenging. For example, as summarized in \cite{sanabria2023audiovisual}, AV16.3 \cite{lathoud2004av16} contains 5-minute annotated sequences and CAV3D \cite{qian2019multi}  contains 14-minute annotated sequences, which is not sufficient for training.
One possible solution to the problem of insufficient data is the teacher-student paradigm, also referred to as knowledge distillation \cite{hinton2015distilling}. 
It adopts a network pre-trained on the desired task (teacher) to automatically extract pseudo-labels from an unlabelled dataset. The pseudo-labels are then used to supervise the training of a new network (student), trained to produce the same results. The student networks are always more light-weighted than the teacher networks.
In audio-visual learning, typically one modality is used to supervise its counterpart.
Under the guidance of visual modality, audio can be used for complicated tasks \cite{zhao2022visually}, including semantic segmentation \cite{irie2019seeing}, depth perception \cite{vasudevan2020semantic}, acoustic scene classification \cite{aytar2016soundnet}, speaker detection and localization \cite{Berghi:2021:mmsp, zhao2022visually} and vehicle localization \cite{gan2019self} \cite{valverde2021there}. In these works a visual teacher network is used to extract positional pseudo-labels to train a multi-channel audio student network. The visual modality can provide beneficial supervision for audio as it has higher spatial accuracy, using color histograms or face detection. In contrast, the audio modality is omnidirectional, presents higher temporal resolution, and does not fail when the speaker is visually occluded. 
\added{Other than teacher-student paradigm, active learning\cite{yuan2023active} and self-supervised learning \cite{yuan2020self} can also be employed to leverage the unlabelled data.}

\section{Methods of Audio-Visual Speaker Tracking}
\label{cls}
\begin{figure}[tbp]
    \centering
    \includegraphics[width=\columnwidth]{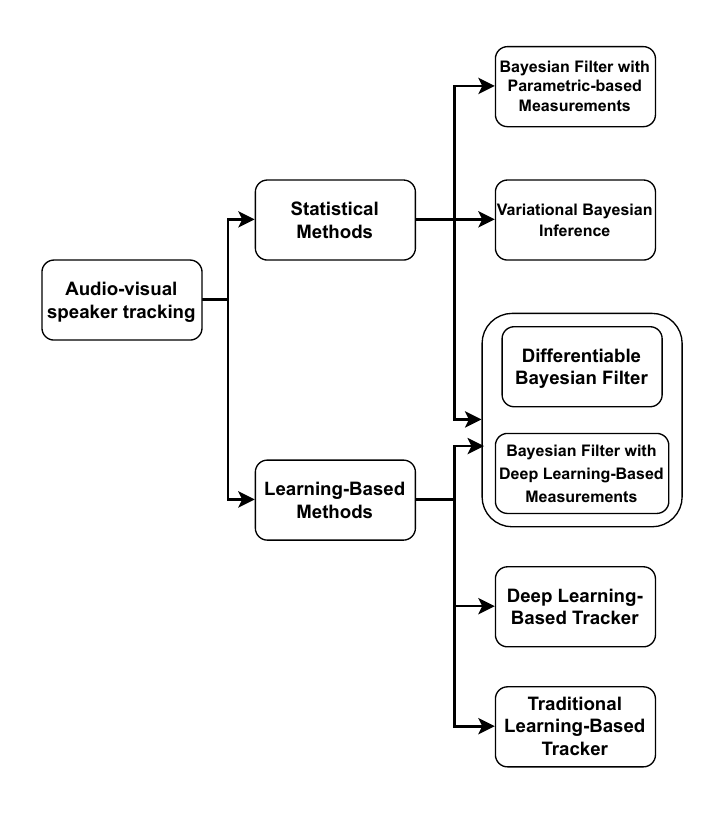}
    \caption{The Classification of Audio-Visual Trackers.}
    \label{classification}
\end{figure}

The classification of the current audio-visual speaker tracker is shown in Fig. \ref{classification}. As most trackers adopt the traditional statistical methods, or use the Bayesian filter with measurements, we start the survey with the Bayesian filter, and then we talk about some emerging techniques such as differentiable Bayesian filters and Transformer-based methods. We summarize the audio-visual speaker trackers in the past few years in Table \ref{overview}.

\subsection{Bayesian Tracking}
\label{Bayesian}

\begin{table*}[htbp]
\caption{Overview of Bayesian Filters}
\centering
\setlength{\tabcolsep}{3mm}{
\begin{tabular}{|l|l|l|l|}
\hline
\textbf{Method}         & \textbf{Propagating Representation}   & \textbf{\begin{tabular}[c]{@{}l@{}}Prediction / \\ Update Function\end{tabular}} & \textbf{No. Targets} \\ \hline
Kalman Filter           & Gaussian Posterior                  & Linear                                                                           & Single               \\ \hline
Extended Kalman Filter  & Gaussian Posterior                  & Nonlinear                                                                        & Single               \\ \hline
Unscented Kalman Filter & Gaussian Posterior                  & Nonlinear                                                                        & Single               \\ \hline
Particle Filter         & SMC Posterior                       & Nonlinear                                                                        & Multiple             \\ \hline
GM-PHD                  & Gaussian Intensity                  & Linear                                                                           & Multiple             \\ \hline
SMC-PHD                 & SMC Intensity                       & Nonlinear                                                                        & Multiple             \\ \hline
GM-Bernoulli Filter     & Gaussian Spatial PDF                & Linear                                                                           & Single               \\ \hline
SMC-Bernoulli Filter    & SMC Spatial PDF                     & Nonlinear                                                                        & Single               \\ \hline
PMBM Filter             & Gaussian Multiple Bernoulli Mixture & Linear                                                                           & Multiple             \\ \hline
\end{tabular}}
\label{filters}
\end{table*}

\begin{figure}[tbp]
    \centering
    \includegraphics[width=0.8\columnwidth]{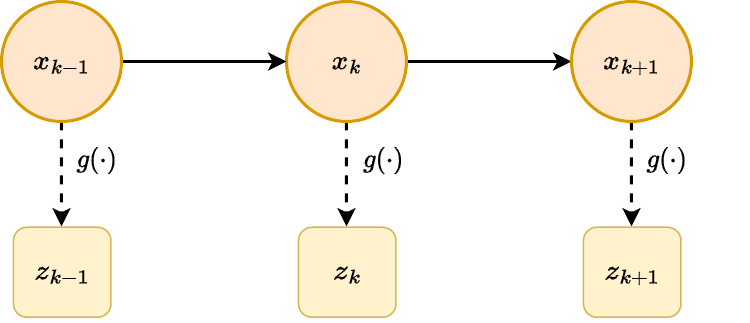}
    \caption{Illustrations of Hidden Markov Model}
    \label{HMM}
\end{figure}

\begin{figure}[tbp]
    \centering
    \includegraphics[width=0.8\columnwidth]{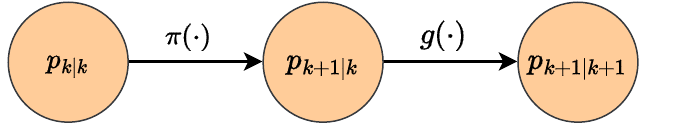}
    \caption{States Updating of Bayesian Filter}
    \label{BF}
\end{figure}

Bayesian based trackers aim to predict target states $\mathbf{x}_{k}$ at time step $k$ recursively given the measurements $\mathbf{z}_{1:k}$. It is assumed that the estimate of target states follows a Markov process of order one, i.e. $\mathbf{x}_{k}$ only depends on $\mathbf{z}_{k}$ and has no relevance with $\mathbf{z}_{1:(k - 1)}$, as shown in Figure \ref{HMM}. Target states are defined as $\mathbf{x} = (x, v_{x}, y, v_{y}) $ in 2D tracking and $\mathbf{x} = (x, v_{x}, y, v_{y}, z, v_{z}) $ in 3D tracking (subscripts $k$ omitted for convenience), where $ (x, y, z) $ is the position of the speaker's mouth and $ (v_{x}, v_{y}, v_{z}) $ is the velocity component. The estimation process contains two steps: prediction and update, as shown in Figure \ref{BF}.

The predicted distribution $p_{k+1 \mid k}\left(\mathbf{x}_{k+1}\right)$ at time step $k + 1$ can be derived by the Chapman Kolmogorov equation:
\begin{equation}
    p_{k+1 \mid k}\left(\mathbf{x}_{k + 1}\right)=\int \pi\left(\mathbf{x}_{k+1} \mid \mathbf{x}_{k}\right) p_{k \mid k}\left(\mathbf{x}_{k}\right) \delta \mathbf{x}_{k}
    \label{eq6}
\end{equation}

\noindent where $ \pi\left(\mathbf{x}_{k+1} \mid \mathbf{x}_{k}\right) $ is the transition density, assumed to have a constant velocity.

The updated distribution at time $ k + 1 $ incorporating measurements $\mathbf{z}_{k + 1}$ can be calculated with the measurement model $ g\left(\mathbf{z}_{k+1} \mid \mathbf{x}_{k+1}\right) $:
\begin{equation}
    p_{k+1 \mid k+1}\left(\mathbf{x}_{k+1}\right)=\frac{g\left(\mathbf{z}_{k+1} \mid \mathbf{x}_{k+1}\right) p_{k+1 \mid k}\left(\mathbf{x}_{k+1}\right)}{\int g\left(\mathbf{z}_{k+1} \mid \mathbf{x}_{k+1}^{\prime}\right) p_{k+1 \mid k}\left(\mathbf{x}_{k+1}^{\prime}\right) \delta \textbf{x}_{\added{k + 1}}^{\prime}}
    \label{eq7}
\end{equation}

We summarize the Bayesian filters in Table \ref{filters}.

\subsubsection{Kalman Filter}
In Kalman filter \cite{kalman1960new}, the posterior distribution $p_{k \mid k}\left(\mathbf{x}_{k}\right)$ and $p_{k+1 \mid k+1}\left(\mathbf{x}_{k+1}\right)$ are Gaussian and the transition process $ \pi\left(\mathbf{x}_{k+1} \mid \mathbf{x}_{k}\right) $ is linear. Kalman filter gives optimal performance with linear Gaussian measurements and has been applied in audio-visual speaker tracking. In  \cite{schymura2020audiovisual} and  \cite{schymura2020dynamic}, a neural network was designed to determine the weights of audio and visual signals dynamically and Kalman filter is used to predict DOA adaptively. In addition to audio-visual speaker tracking, Kalman filter has been used in Multiple Object Tracking (MOT). In \cite{bewley2016simple}, \added{a simple online and real-time tracking algorithm (SORT)} was proposed which combined Kalman filter with Hungarian algorithm \cite{kuhn1955hungarian} for motion prediction and data association. In \cite{wojke2017simple}, the same estimation model in \cite{bewley2016simple} was employed and the appearance information of objects was integrated to improve the performance. In \cite{wojke2017simple} and \cite{bewley2016simple}, the object motion is considered linear due to the high frame per second (FPS) of the camera.

Kalman filter is not applicable to non-linear measurements due to its use of linear and Gaussian models. Extensions of Kalman filter such as extended Kalman filter (EKF) \cite{simon2006optimal} and unscented Kalman filter (UKF) \cite{wan2001unscented} can handle non-linear measurements. EKF uses a local linearization by utilizing the first term in a Taylor expansion of the nonlinear function to deal with the non-linearity. UKF mitigates the problem of non-linearity by approximating the state distribution by a set of points. In \cite{d2012person}, EKF was employed to process audio and visual measurements separately, and to fuse the two estimations at the decision level to obtain the final results.

\added{In summary, Kalman filter is applicable in linear and Gaussian environment. It is suitable for single speaker tracking where the transition model and update model are known.}

\subsubsection{Particle Filter}

Particle filter is a sequential Monte Carlo (SMC) algorithm and has better performance than EKF and UKF. However, PF suffers from a weight degeneracy problem if only a few particles contribute to the state estimation after several iterations. A resampling algorithm is proposed to mitigate the effect of degeneracy \cite{arulampalam2002tutorial}. The main idea of resampling is to duplicate important particles and discard unimportant ones. The resampling methods include multinomial resampling, residual resampling \cite{liu1998sequential}, stratified resampling and systematic resampling \cite{kitagawa1996monte}. However, resampling may create many repeated particles, which lowers their diversity. The traditional intelligent optimization algorithms can be combined with particle filters to maintain the diversity of particles such as particle swarm
optimization algorithm \cite{fang2007particle} and firefly algorithm \cite{tian2019new, tian2022intelligent}.

In the general framework of a particle filter, there are four fundamental steps: initialization, prediction, update, and resampling. Particles, denoted as $ \boldsymbol{p}^{(i)}_{k} $ at time step $k$, are employed to represent the state of an object, with $i$ serving as the particle index.
In the initial stage, all particles share the same weight, which is uniformly distributed as $w^{(i)}_{0} = \frac{1}{N}$ with $N$ denoting the number of particles. In the prediction step, particle states are advanced by:
\begin{equation}
\boldsymbol{p}_{k}^{(i)}=\boldsymbol{F} \boldsymbol{p}_{k-1}^{(i)}+\boldsymbol{q}_{k}^{(i)}
\end{equation}

\noindent where $\boldsymbol{F}$ represents the prediction matrix and $\boldsymbol{q}_{k}^{(i)}$ denotes the Gaussian noise $\mathcal{N} (0, \boldsymbol{Q}^2)$. \added{$\boldsymbol{Q}$ is the covariance and is pre-defined as a hyper-parameter in practice.} 

In the update step, particles' weights are adjusted by the measurement model with measurements $\boldsymbol{z}_{k}$. 
\begin{equation}
    \omega_{k}^{(i)} \propto g\left(\boldsymbol{z}_{k} \mid \boldsymbol{p}_{k}^{(i)}\right)
\end{equation}

Subsequently, speakers' states can be updated by the weighted average over the particle states:
\begin{equation}
    \boldsymbol{x}_{k}=\sum_{i=1}^{N} \omega_{k}^{(i)} \boldsymbol{p}_{k}^{(i)}
\end{equation}
The final step in the process is resampling, where particles with large weights are preserved and copied for the subsequent time step, while particles with small weights are removed. This step ensures that the particle set remains representative of the evolving state distribution.

\added{In summary, particle filter can be applied in non-linear and non-Gaussian scenario. It can be used for multi-speaker tracking and the particle representation is suitable for multi-modal fusion. The drawback is the high computational cost due to a large number of particles. And the model performance is sensitive to the particle initialization. }

\subsubsection{RFS}

Both the KF and PF algorithms assume that the number of tracking targets is known and fixed. If the number of targets varies with time, these algorithms may not work well. Random Finite Set (RFS) is proposed to model the evolution of objects with unfixed quantities. RFS depicts the process of motion, birth and death of targets.

From the perspective of RFS, for a target with the state $\boldsymbol{x}_{k}$ at time $k$, it has the surviving possibility $P_{S}$ to exist at time $k + 1$ and evolves to the state $\boldsymbol{x}_{k + 1}$ with the transition function $ \pi\left(\boldsymbol{x}_{k+1} \mid \boldsymbol{x}_{k}\right) $, or has the possibility $1-P_{S}$ to die. At the same time, new targets may appear. The new targets come from two parts, namely, the targets spawned from the existing targets, and the born targets that are independent from the existing targets. Therefore, The multi-target states at time $k + 1$ are from three aspects: surviving targets at time $k + 1$, the spawned targets from time $k$ and new born targets at time $k + 1$.

The state distribution of measurements $z_{k}$ at time $k$ also follows RFS. The target $x_{k}$ has the possibility $P_{D}$ to be detected or has the possibility $1 - P_{D}$ to be missed. Apart from the measurements from the targets, the audio or visual sensors may generate clutter such as false positive detection.

\subsubsection{PHD Filter}
PHD filter is one of the RFS-based methods. PHD filter is short for Probability Hypothesis Density  Filter, which transmits the first-order moment of the posterior density to lower the computational complexity. The first-order moment is also called the intensity $v$, whose integral is the estimated number of speakers. The PHD filter contains the prediction and update steps. The prediction is expressed as follows:
\begin{equation}
    \begin{aligned}
v_{k+1 \mid k}\left(\mathbf{x}_{k+1}\right)&=\gamma_{k+1}\left(\mathbf{x}_{k+1}\right) 
+\\&\int 
p_{S}  \pi_{k+1 \mid k}\left(\mathbf{x}_{k+1} \mid \mathbf{x}_{k}\right) v_{k \mid k}\left(\mathbf{x}_{k}\right) d \mathbf{x}_{k}
+\\&\int \beta_{k+1 \mid k}\left(\mathbf{x}_{k+1} \mid \mathbf{x}_{k}\right)
 v_{k \mid k}\left(\mathbf{x}_{k}\right) d \mathbf{x}_{k}
\end{aligned}
\end{equation}

\noindent where $v_{k+1 \mid k}\left(\mathbf{x}_{k+1}\right)$ is the new speakers birth intensity and $\pi_{k+1 \mid k}\left(\mathbf{x}_{k+1} \mid \mathbf{x}_{k}\right)$ is the states transition function defined as before. $\beta_{k \mid k-1}\left(\mathbf{x}_{k} \mid \mathbf{x}_{k-1}\right)$ is the intensity of the speakers spawned from $\boldsymbol{x}_{k}$. The update process is expressed as follows:
\begin{equation}
    \begin{aligned}
&v_{k+1 \mid k+1}\left(\mathbf{x}_{k+1}\right) =\left(1-p_{D}\right) v_{k+1 \mid k}\left(\mathbf{x}_{k+1}\right) 
+ \\&\sum_{\mathbf{z}_{k+1} \in \mathcal{Z}_{k+1}} \frac{p_{D} g_{k+1}\left(\mathbf{z}_{k+1} \mid \mathbf{x}_{k+1}\right) v_{k+1 \mid k}\left(\mathbf{x}_{k+1}\right)}{\kappa_{k+1}+\int p_{D} g_{k+1}\left(\mathbf{z}_{k+1} \mid \mathbf{x}_{k+1}\right) v_{k+1 \mid k}\left(\mathbf{x}_{k+1}\right)}
\end{aligned}
\end{equation}

where $P_{D}$ is the detection probability, $g_{k+1}\left(\mathbf{z}_{k+1} \mid \mathbf{x}_{k+1}\right)$ is the measurement likelihood function given measurements $z_{k+1}$, and $\kappa_{k+1}$ is the clutter intensity.

The PHD filter has no closed-form solutions and only has two numerical solutions: the Gaussian mixture form (GM-PHD) \cite{vo2006gaussian} and the SMC form (SMC-PHD) \cite{vo2005sequential}. The latter form has been widely used as it does not require the linear Gaussian assumption. Audio-visual SMC-PHD \cite{liu2017particle} filter is proposed for multiple speakers tracking using both audio and visual information. Audio information is used to relocate particles combined with particle flow \cite{daum2007nonlinear} and to detect new speakers while the visual information is used to update the particle weights.

\added{RFS-based filters are suitable for the scenario where the speakers move in and out, and the number of speakers is changing frequently. However, it is computationally expensive when there is a large number of speakers.}

\added{Apart from the mentioned PHD filter, there are also other RFS-based filters such as multi-target multi-Bernoulli filter \cite{vo2008cardinality} and poisson multi-Bernoulli mixture \cite{garcia2018poisson}. }

\subsection{Variational Bayesian Inference}
In \cite{ban2019variational}, the problem of audio-visual speaker tracking is formulated as a temporal graphical model with latent variables. The objective is to maximize the posterior joint distribution of the latent variables based on the audio and visual measurements. Variational expectation maximization is used to deal with the intractable estimation. First, the expectation with regard to the latent variables is calculated for the posterior likelihood. Then the posterior likelihood is maximized to estimate the model parameters.

\subsection{Tracking with Deep Learning}
\label{track_deep_learning}
Most of the existing works for audio-visual speaker tracking employ Bayesian filters for data association and track management but very few works tried to use deep learning. The main reason is that the audio-visual speaker tracking datasets are not large enough to train a tracking network. 

Several works on similar tasks have tried to use deep learning frameworks for tracking. In \cite{zhang2019robust}, an end-to-end framework is proposed for tracking on KITTI Tracking Benchmark \cite{geiger2012we}, in which images of the visual and LiDAR modalities are provided. The proposed tracking framework ensembles feature extraction, data association and track management. In addition, in the MOT challenge, there are some works \cite{ondruska2016deep} \cite{milan2017online} using \added{Recurrent Neural Network (RNN)} for a unified end-to-end tracking framework. Trackformer is proposed in \cite{meinhardt2022trackformer}, which is based on Transformer. The encoder takes the image as input and the decoder takes object queries as input. Each query corresponds to a potential object. The output of the decoder which indicates the appearance of the object will be delivered to the next time step as new queries. TransTrack \cite{sun2020transtrack} leverages two sets of queries. The one is object query, acting as an object detector. The other is track query, which associates objects in the current frame with those in previous frames. \added{Intersection over Union (IOU)} matching is employed to associate the detected objects with tracklets. ByteTrack is proposed in \cite{zhang2022bytetrack}. Different from previous trackers, ByteTrack not only associates high confidence bounding boxes but the low confidence boxes. In MotionTrack \cite{qin2023motiontrack}, an interaction module is designed for short-term association and a refind module is designed for long-term matching, which achieves a good performance under dense crowds and occlusions. \added{In \cite{yang2023hard}, buffered IoU matching is propose. Buffered IoU enlarges the area of bounding box. The strategy is to match the alive tracks with a small buffer size and match the unmatched tracks with a large size, which mitigates the problem of irregular motion.} In other survey papers \cite{xu2019deep, wang2022recent} with regard to MOT, more deep learning-based trackers are provided. In other tracking tasks such as 3D MOT \cite{willes2023intertrack} and infrared tracking \cite{geng2022person}, deep learning-based are also widely adopted.

More recently, AVRI \cite{qian2022audio} dataset has been proposed for audio-visual speaker localization and tracking. It contains more than nine hours of audio-visual data and enables the employment of deep learning techniques. However, AVRI can only be used for single-speaker tracking. Overall, using deep learning for audio-visual speaker tracking remains an open and challenging problem. The problem of the lack of large amounts of datasets needs to be overcome. 
\added{In summary, deep learning methods are suitable for the scenario where the dataset size is large enough to train the measurement model and the tracker.}

\added{In addition to the trackers using deep learning-based backbones, Bayesian trackers with deep learning-based measurements can also be classified as deep learning-based methods, which can be found in }Table \ref{overview}.

\subsection{Tracking with Traditional Learning Methods}
\added{Apart from deep learning methods, there are also traditional learning-based methods used in tracking. In \cite{barnard2014robust}, dictionary learning is used to model the appearance of the speakers. Then support vector machine (SVM) is used for classification of head and background using image histograms of learned dictionary. Finally the estimated likelihood by SVM is used in PF for tracking.} 

\subsection{Differentiable Bayesian Filter}
\subsubsection{Learnable Prediction and Updating}
Bayesian filters can also be designed to be differentiable so that the motion model in the prediction stage and the measurement-correct model in the update stage can be trained and optimized end-to-end through deep learning models. The transition model (Equation \ref{eq6}) and update model (Equation \ref{eq7}) can be replaced by deep learning modules. 
\begin{equation}
    p_{k+1 \mid k}\left(\mathbf{x}_{k + 1}\right)= T(p_{k \mid k}\left(\mathbf{x}_{k}\right)) 
\end{equation}
\begin{equation}
    p_{k+1 \mid k+1}\left(\mathbf{x}_{k+1}\right)=F(\mathbf{z}_{k+1},  p_{k+1 \mid k}\left(\mathbf{x}_{k+1}\right))
\end{equation}
\noindent where $T(\cdot)$ and $F(\cdot)$ denote the learnable transition and update model, which can be fed forward layers, convolutional neural networks, recurrent neural networks or Transformer modules. Compared to traditional Bayesian filters, the learnable models are more flexible to adapt to different scenarios. 
\subsubsection{Soft-Resampling}
In particle filter, resampling is required to select the important particles and discard the low-weight particles. However, the resampling operation is not differentiable. To solve this problem, soft-resampling is proposed, which introduces a unified distribution $ u(\cdot)$ mixed with updated particle distributions $p(\cdot)$. 
\begin{equation}
    q(i) = \alpha p(i) + (1 - \alpha) u(i)
\end{equation}
where $0 < \alpha < 1$ is the hyperparameter. New particles are sampled from the new distributions $q(\cdot)$ instead of $p(\cdot)$. The weights of new particles are computed as follows:
\begin{equation}
    \hat{w}_t^{k}=\frac{p(k)}{q(k)}=\frac{w_t^k}{\alpha w_t^k+(1-\alpha) 1 / K}
\end{equation}
With soft-resampling, particle filter can be designed to an end-to-end differentiable architecture.

\subsubsection{Summarizations of Current Differentiable Bayesian Trackers}
We summarize the differentiable Bayesian filters in Table \ref{differentiable}. In \cite{haarnoja2016backprop}, the differentiable Kalman filter is proposed. In \cite{kim2022review}, a review of deep learning methods combining the Kalman filter is presented. In \cite{jonschkowski2018differentiable}, a differentiable particle filter was proposed and applied in visual odometry task \cite{geiger2013vision}. The proposed differentiable model excludes the resampling part, which may ignore the effects of estimations in the last time step on the estimations in the current time step. In \cite{karkus2018particle}, a differentiable particle filter was applied in robot visual localization but with a differentiable soft-resampling step, which is beneficial for future estimations. These works show that the Bayesian algorithm priors (i.e., the prediction and the update) enable the explainability of the network, benefit the training process and lead to superior performance compared to pure deep learning models such as \added{long short-term memory (LSTM)}. In \cite{zhu2020towards}, a novel differentiable resampling technique is designed based on the weighted multi-head attention, which is superior to the conventional resampling methods such as soft resampling and systematic resampling. 

\added{In summary, differentiable Bayesian filter can be applied in the scenarios where traditional Bayesian filter does not perform well due to complex motion model or the need for data-driven adaptability.}

\begin{table}[htbp]
\caption{Summarization of Differentiable Bayesian Filters. (\textbf{FC} denotes the fully connected layer, \textbf{TRN} denotes Transformer.)}
\centering
\setlength{\tabcolsep}{3mm}{
\begin{tabular}{|l|l|l|l|}

\hline
\textbf{Ref} & \textbf{Backbone} & \textbf{Bayesian Filter} & \multicolumn{1}{c|}{\textbf{Tasks}} \\ \hline
\cite{haarnoja2016backprop}           & FC                & Kalman Filter            & State Estimation       \\ \hline
\cite{schymura2020dynamic}           & FC                & Kalman Filter            & Speaker Tracking       \\ \hline
\cite{karkus2018particle}           & CNN               & Particle Filter          & Visual Localization                 \\ \hline
\cite{jonschkowski2018differentiable}           & CNN               & Particle Filter          & Global Localization                 \\ \hline
\cite{ma2020particle}           & RNN               & Particle Filter          & \begin{tabular}[c]{@{}l@{}}Robot Localization,\\ Sequence Prediction\end{tabular}                 \\ \hline
\cite{zhu2020towards}           & TRN               & Particle Filter          & Resampling                 \\ \hline
\cite{zhang2023kalmannet}           & RNN               & Kalman Filter          & Echo Cancellation                 \\ \hline
\cite{jouaber2021nnakf}           & RNN               & Kalman Filter          & Noise Estimation                 \\ \hline
\cite{zhao2024attention}           & TRN               & Particle Filter          & Speaker Tracking                 \\ \hline
\end{tabular}}
\label{differentiable}
\end{table}

\subsection{Comparison between Bayesian Filter and Deep Learning-Based Tracker}

\added{Bayesian filter is based on probabilistic model and has Interpretability. It is well-performed and robust in environment where the assumptions of transition and update models hold. However, in practice, the performance degrades in complicated scenarios when the assumptions are not met. Besides, some hyper-parameters in Bayesian filters need to be tuned mannually.}

\added{Deep learning-based methods can be robust in complicated scenarios through training. But it requires abundant training data for measurement extraction and tracking management.}



\section{Other Modules in Tracking Systems}
\subsection{Audio-Visual Fusion}
\label{AVF}
\added{Audio-visual fusion aims to project features from different modalities into the same space so that they can contribute to the calculation of speaker states.}

The methods for audio-visual fusion can be classified into three types: early fusion, late fusion and intermediate fusion \cite{michelsanti2021overview}. 

Early fusion methods combine the audio and visual features before the Bayesian inference. Few works use this method as the feature representation between the audio and visual modalities is inherently different and combining the heterogeneous information at an early stage is a challenging problem. In \cite{qian2021multi}, GCC-PHAT and simulated visual features are encoded concurrently to predict DOA estimations.

Late fusion makes the final decision by combining the decisions from an audio tracker and a visual tracker. There are several works using late fusion. Kalman filter was used in \cite{d2012person} to process audio and visual signals separately, and then Gaussian distribution is used and fused to get the object states. In \cite{qian20173d}, GCF and face detectors were used to get two estimated positions, and PF was employed to fuse the two decisions. 

Intermediate fusion allows the audio and visual signals to interact with each other before making decisions and is the most widely used fusion method. In \cite{qian2019multi}, the height estimation by visual modality was employed to assist the GCF calculation. In \cite{liu20193d}, two particle filters were used to process audio and visual streams independently, and the confidence of audio and visual modalities is leveraged to dynamically adjust the weights of particles. In \cite{li2021multi}, a multi-modal perception attention network was proposed with a self-supervised cross-modal strategy to determine the importance of audio and visual measurements. In \cite{wu2021binaural}, binaural audios and visual frames are encoded separately and fused through ConvLSTM before the decoder for sound source localization.

\subsection{Position Conversion between 2D and 3D}
\added{Position conversion aims to align the measurements in different coordinates. For example, measurements derived by face detector are in 2D coordinates and measurements derived by GCF are in 3D coordinates. Measurements from different coordinates need to be aligned before fusion.}

\begin{figure}[tbp]
    \centering
    \includegraphics[width=0.9\columnwidth]{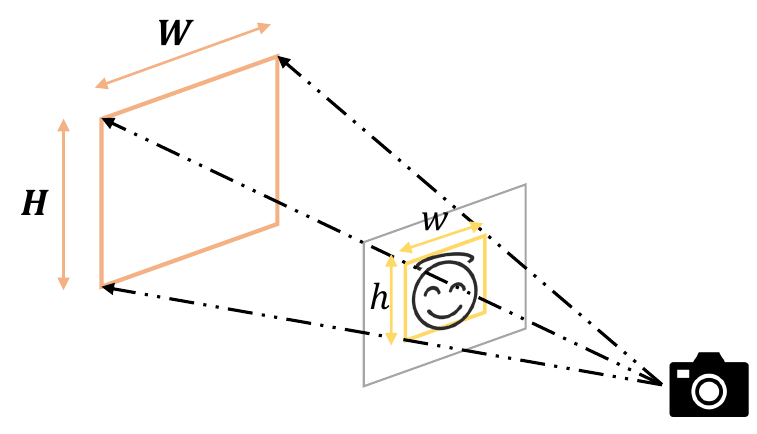}
    \caption{The Camera Projection Model}
    \label{camera}
\end{figure}

The 2D position in the image plane and the 3D position in the world coordinates can be converted to each other through the camera projection model \cite{hartley2003multiple}, as shown in Figure \ref{camera}. 
The 2D position $\mathbf{o}$ derived by face detection can be converted to 3D position $\mathbf{O}$:
\begin{equation}
    \mathbf{O} = \Phi(\mathbf{o}; w, h, W, H)
\end{equation}
where $(w, h)$ is the width and height of the face bounding boxes. $(W, H)$ is the preset width and height of the face bounding box in the 3D space.

Similarly, the 3D position obtained by the sound source localization algorithm can be converted to a 2D position:
\begin{equation}
    \mathbf{o} = \Psi(\mathbf{O}; W, H)
\end{equation}
\added{Both $\Phi$ and $\Psi$ are projection operators.}

\subsection{Data Association}
\label{DataA}
In the scenario of multiple speaker tracking and single target tracking with clutter and false alarms, data association is needed, which contains two aspects of audio-visual speaker tracking. The one is to associate audio measurements with corresponding visual measurements for audio-visual fusion, which is discussed in \added{Section V.\ref{AVF}}. The other is to associate the fused measurements to the existing tracks, clutter and the new tracks. 

The Nearest Neighbor (NN) algorithm is the simplest data association method, which associates the closest measurements with the target. NN regards the data association as linear programming and minimizes the association cost globally. While NN is easy to associate clutter or false alarms with targets and deviates from the tracks.
JPDA \cite{rezatofighi2015joint} is short for Joint Probabilistic Data Association, which assumes that each measurement originates from clutter or targets and each target can only generate one measurement. Each target may have multiple effective measurements. JPDA calculates the joint probability of targets associated with different effective measurements. The drawback of JPDA is that it requires the prior of the number of targets and fails in the scenario of missing targets.
MHT \cite{kim2015multiple} (Multiple Hypothesis Tracking) maintains an association hypothesis tree and calculates possibilities of all association hypothesis branches. A new measurement can be associated with an old hypothesis, can start a new hypothesis and can be a false alarm. The computational cost of MHT is increasing exponentially with the number of measurements. To lower the computational cost, the hypothesis with low possibility can be deleted and similar hypotheses can be merged.

\renewcommand\arraystretch{1.2}

\begin{table*}[htbp]
\centering
\caption{Multi-modal Datasets. No. Mic denotes the number of microphones over all microphone arrays; SR denotes the sampling rate with the unit of kHz; CA (Circular microphone Arrays) denotes whether the microphone array is in the circular shape; No. Cam denotes the number of cameras; Fps denotes frame per second; Co (Co-located) denotes whether the multi-modal sensors are co-located; Cal (Calibration information) denotes whether the camera calibration information is available; VAD (Voice activity Detectors) denotes whether the speakers' states are available; Type denotes the formats of the annotation; No. Spk denotes the maximum number of speakers in one frame and `-' denotes `not applicable'.}
\setlength{\tabcolsep}{0.8mm}{
\begin{tabular}{|l|ccc|ccc|c|ccl|c|}
\hline
\multicolumn{1}{|c|}{\multirow{2}{*}{Dataset}} & \multicolumn{3}{c|}{Audio}                                                             & \multicolumn{3}{c|}{Video}                                                                         & \multicolumn{1}{l|}{\multirow{2}{*}{Co}} & \multicolumn{3}{c|}{Annotation}                                                                                                                                                                                           & \multicolumn{1}{l|}{\multirow{2}{*}{No. Spk}} \\ \cline{2-7} \cline{9-11} 

\multicolumn{1}{|c|}{}                         & \multicolumn{1}{l|}{No. Mic} & \multicolumn{1}{l|}{SR (kHz)} & \multicolumn{1}{l|}{CA} & \multicolumn{1}{l|}{No. Cam} & \multicolumn{1}{l|}{Resolution (pixels)} & \multicolumn{1}{l|}{Fps} & \multicolumn{1}{l|}{}                    & \multicolumn{1}{l|}{Cal} & \multicolumn{1}{l|}{VAD} & \multicolumn{1}{c|}{Type}                                                                                                                                           & \multicolumn{1}{l|}{}                         \\ \hline

AVTRACK-1 \cite{gebru2015tracking}                                      & \multicolumn{1}{c|}{\num{4}}       & \multicolumn{1}{c|}{\num{44.1}}     & -                       & \multicolumn{1}{c|}{2}       & \multicolumn{1}{c|}{640 $\times$ 480}           & 25                       & \checkmark                                        & \multicolumn{1}{c|}{-}   & \multicolumn{1}{c|}{\checkmark}   & \begin{tabular}[c]{@{}l@{}}Active speaker(s) bounding box, \\  upper-body region bounding box\end{tabular}                                                          & 2                                             \\ \hline
AVASM \cite{deleforge2015co}                                         & \multicolumn{1}{c|}{2}       & \multicolumn{1}{c|}{44.1}     & -                       & \multicolumn{1}{c|}{2}       & \multicolumn{1}{c|}{640 $\times$ 480}           & -                        & \checkmark                                        & \multicolumn{1}{c|}{-}   & \multicolumn{1}{c|}{-}   & 2D coordinates of a loud-speaker                                                                                                                                    & 2                                             \\ \hline
AVDIAR \cite{gebru2017audio}                                        & \multicolumn{1}{c|}{6}       & \multicolumn{1}{c|}{48}       & -                       & \multicolumn{1}{c|}{2}       & \multicolumn{1}{c|}{1920 $\times$ 1200}         & 25                       & \checkmark                                        & \multicolumn{1}{c|}{\checkmark}   & \multicolumn{1}{c|}{\checkmark}   & \begin{tabular}[c]{@{}l@{}}2D coordinates of the head and \\ upper-body\end{tabular}                                                                                & 4                                             \\ \hline
RAVEL \cite{alameda2013ravel}                                         & \multicolumn{1}{c|}{4}       & \multicolumn{1}{c|}{48}       & -                       & \multicolumn{1}{c|}{2}       & \multicolumn{1}{c|}{1024 $\times$ 768}          & 15                       & \checkmark                                       & \multicolumn{1}{c|}{\checkmark}   & \multicolumn{1}{c|}{\checkmark}   & \begin{tabular}[c]{@{}l@{}}Both 2D and 3D coordinates of \\ actors'  positions\end{tabular}                                                                         & 5                                             \\ \hline
CAVA \cite{arnaud2008cava}                                          & \multicolumn{1}{c|}{2}       & \multicolumn{1}{c|}{44.1}     & -                       & \multicolumn{1}{c|}{2}       & \multicolumn{1}{c|}{1024 $\times$ 768}          & 25                       & \checkmark                                        & \multicolumn{1}{c|}{\checkmark}   & \multicolumn{1}{c|}{-}   & 3D trajectory of head tracking                                                                                                                                      & 5                                             \\ \hline
SPEVI \cite{taj2007surveillance}                                         & \multicolumn{1}{c|}{2}       & \multicolumn{1}{c|}{44.1}     & -                       & \multicolumn{1}{c|}{1}       & \multicolumn{1}{c|}{360 $\times$ 288}           & 25                       & \checkmark                                        & \multicolumn{1}{c|}{-}   & \multicolumn{1}{c|}{-}   & Face bound boxes                                                                                                                                                    & 2                                             \\ \hline
AMI  \cite{carletta2006announcing}                                          & \multicolumn{1}{c|}{\added{14+}}       & \multicolumn{1}{c|}{48}       & \checkmark                       & \multicolumn{1}{c|}{\added{2+}}       & \multicolumn{1}{c|}{720 $\times$ 576}           & 25                       & -                                        & \multicolumn{1}{c|}{-}   & \multicolumn{1}{c|}{\checkmark}   & \begin{tabular}[c]{@{}l@{}}Occlusion status,  head, hand \\ and face positions\end{tabular}                                                                         & \added{5}                                             \\ \hline
CHIL \cite{mostefa2007chil}                                          & \multicolumn{1}{c|}{88}      & \multicolumn{1}{c|}{44.1}     & -                       & \multicolumn{1}{c|}{5}       & \multicolumn{1}{c|}{1024 $\times$ 768}          & 30                       & -                                        & \multicolumn{1}{c|}{\checkmark}    & \multicolumn{1}{c|}{\checkmark}    & \begin{tabular}[c]{@{}l@{}}Face, head, eyes and \\ nose positions\end{tabular}                                                                                      & 5                                             \\ \hline
AV16.3 \cite{lathoud2004av16}                                        & \multicolumn{1}{c|}{16}      & \multicolumn{1}{c|}{16}       & \checkmark                       & \multicolumn{1}{c|}{3}       & \multicolumn{1}{c|}{360 $\times$ 288}           & 25                       & -                                        & \multicolumn{1}{c|}{\checkmark}   & \multicolumn{1}{c|}{\checkmark}   & \begin{tabular}[c]{@{}l@{}}Both 2D and 3D face and \\ head positions\end{tabular}                                                                                   & 3                                             \\ \hline
CAV3D \cite{qian2019multi}                                         & \multicolumn{1}{c|}{8}       & \multicolumn{1}{c|}{96}       & \checkmark                       & \multicolumn{1}{c|}{1}       & \multicolumn{1}{c|}{1024 $\times$ 768}          & 15                       & \checkmark                                        & \multicolumn{1}{c|}{\checkmark}   & \multicolumn{1}{c|}{\checkmark}   & Mouth positions                                                                                                                                                     & 3                                             \\ \hline
CLEAR \cite{ooi2007evaluation}                                         & \multicolumn{1}{c|}{14+}    & \multicolumn{1}{c|}{44.1}     & -                       & \multicolumn{1}{c|}{4}       & \multicolumn{1}{c|}{1024 $\times$ 768}          & 30                       & \checkmark                                        & \multicolumn{1}{c|}{\checkmark}   & \multicolumn{1}{c|}{\checkmark}   & \begin{tabular}[c]{@{}l@{}}Both 2D and 3D head \\ locations, face bound boxes\end{tabular}                                                                          & 8                                             \\ \hline
S3A \cite{liu2017multiple}                                           & \multicolumn{1}{c|}{2}       & \multicolumn{1}{c|}{44.1}     & -                       & \multicolumn{1}{c|}{1}       & \multicolumn{1}{c|}{-}                   & 30                       & -                                        & \multicolumn{1}{c|}{-}   & \multicolumn{1}{c|}{-}   & \begin{tabular}[c]{@{}l@{}}No visual infomation is provided \\ but depth information is provided\end{tabular}                                                       & -                                             \\ \hline
TragicTalkers\,\cite{Berghi:2022:TragicTalkers}                                            & \multicolumn{1}{c|}{38}      & \multicolumn{1}{c|}{48}       & -                       & \multicolumn{1}{c|}{22}      & \multicolumn{1}{c|}{2448 $\times$ 2048}         & 30                       & \checkmark                                        & \multicolumn{1}{c|}{\checkmark}   & \multicolumn{1}{c|}{\checkmark}   & 3D mouth positions and pesudo labels                                                                                                                                                 & 2                                             \\ \hline
AVOT \cite{wilson2020avot}                                          & \multicolumn{1}{c|}{-}       & \multicolumn{1}{c|}{44.1}     & -                       & \multicolumn{1}{c|}{-}       & \multicolumn{1}{c|}{800 $\times$ 600}           & -                        & -                                        & \multicolumn{1}{c|}{-}   & \multicolumn{1}{c|}{-}   & \begin{tabular}[c]{@{}l@{}}2D positions of tabletoped \\ sized objects\end{tabular}                                                                                 & -                                             \\ \hline
SSLR \cite{he2018deep}                                       & \multicolumn{1}{c|}{4}       & \multicolumn{1}{c|}{48}       & -                       & \multicolumn{1}{c|}{-}       & \multicolumn{1}{c|}{-}                   & -                        & -                                        & \multicolumn{1}{c|}{-}   & \multicolumn{1}{c|}{\checkmark}   & 3D positions of sound sources                                                                                                                                       & 2                                             \\ \hline
MAVD \cite{valverde2021there}                                          & \multicolumn{1}{c|}{8}       & \multicolumn{1}{c|}{44.1}     & \checkmark                       & \multicolumn{1}{c|}{2}       & \multicolumn{1}{c|}{1920 $\times$ 650}          & -                        & \checkmark                                        & \multicolumn{1}{c|}{-}   & \multicolumn{1}{c|}{-}   & \begin{tabular}[c]{@{}l@{}}No ground truth available. In addition \\ to audio and visual modalities, thermal \\ and depth modalities are also provided\end{tabular} & -                                             \\ \hline
AVIAD \cite{perez2020audio}                                         & \multicolumn{1}{c|}{128}     & \multicolumn{1}{c|}{12}       & -                       & \multicolumn{1}{c|}{1}       & \multicolumn{1}{c|}{640 $\times$ 480}           & -                        & \checkmark                                        & \multicolumn{1}{c|}{-}   & \multicolumn{1}{c|}{-}   & People's actions                                                                                                                                                    & 1                                             \\ \hline
AVRI \cite{qian2022audio}                                         & \multicolumn{1}{c|}{4}     & \multicolumn{1}{c|}{\added{16}}       & -                       & \multicolumn{1}{c|}{1}       & \multicolumn{1}{c|}{960 $\times$ 540}           & -                        & \checkmark                                        & \multicolumn{1}{c|}{-}   & \multicolumn{1}{c|}{-}   & Azimuth (Direction of Arrival)                                                                                                                               & 1                                             \\ \hline

EasyCom \cite{donley2021easycom}                                         & \multicolumn{1}{c|}{4+}     & \multicolumn{1}{c|}{48}       & \checkmark                       & \multicolumn{1}{c|}{1}       & \multicolumn{1}{c|}{1920 $\times$ 1080}           & 20                        & \checkmark                                        & \multicolumn{1}{c|}{\checkmark}   & \multicolumn{1}{c|}{\checkmark}   & 3D positions and rotations                                                                                                                     & 1                                             \\ \hline
\end{tabular}}
\label{t1}
\end{table*}

\section{Datasets}
\label{dataset}
Most datasets for audio-visual speaker tracking are recorded by microphone arrays and cameras. The microphone array can be circular, planar, T-shaped or in other shapes. The audio-visual datasets can be classified as co-located or spatially distributed depending on whether the multi-modal sensors are co-located or not. Most datasets provide the recording timing sequences used to synchronize the multi-modal sequences. The annotations contain the camera calibration information, voice activity detectors and ground truth positions. Camera calibration information is usually used to project 2D coordinates to 3D coordinates or project in reverse \cite{hartley2003multiple}. Voice activity detectors denote whether the speakers are talking over frames. Ground truth positions are often the face bounding boxes and mouth positions. In \cite{qian2019multi}, some audio-visual datasets were summarized for speaker tracking. We give a more thorough review of the commonly used datasets in Table \ref{t1}. In addition to the audio-visual speakers tracking datasets, we also list some multi-modal datasets for objects such as vehicle \cite{valverde2021there} and small objects \cite{wilson2020avot}. Besides, datasets containing other modalities such as depth \cite{liu2017multiple} and thermal maps \cite{valverde2021there} are included. We also introduce some influential datasets in detail.

\subsection{AV16.3 Dataset}
AV16.3 \cite{lathoud2004av16} dataset is widely employed for evaluating speaker localization and tracking systems. AV16.3 dataset was captured using two circular microphone arrays and three cameras. For audio, each microphone array has 8 microphones arranged in a circular geometry, recording at a 16 kHz sampling rate. For video, three synchronized cameras capture images at 25 frames per second. The two microphone arrays are located \num{0.8} meters apart at a table. Within the AV16.3 dataset, speakers in the recording space engage in various activities, including sitting statically, standing statically, or walking near the table. The duration of most sequences range from 20 to 60 seconds, although there are also longer sequences that extend beyond three minutes. AV16.3 dataset contains over 40 different audio-visual sequences. However, only a small subset of these sequences have annotated ground truth labels. Sequences 08, 11, 12, 19 and 20 are often used for the evaluation of single-speaker tracking while sequences 24, 25 and 30 are often used for the evaluation of multiple-speaker tracking. These particular sequences are challenging due to occurrences of occlusions and instances where speakers are not facing the cameras, adding complexity to the tracking evaluation process.

\subsection{CAV3D Dataset}

CAV3D is a dataset recorded by Co-located Audio-Visual sensors for 3D tracking. It is recorded in a  $4.66 \times 5.95 \times 4.5$ room with an eight-microphone circular array with a sample rate of 96kHz and a camera with \added{15} fps. This dataset contains nine single-speaker sequences and 11 multiple-speaker sequences. Compared to the AV16.3 dataset, scenarios in the CAV3D dataset are more challenging as it contains more frames where speakers are occluded by each other and speakers are outside the field of view of the camera. 

\subsection{AVRI Dataset}
Different from the previous dataset, AVRI (Audio-Visual Robotic Interface) has around 9 hours, which provides sufficient data for training a neural network. ReSpeaker microphone array is used to record the multi-channel audio with 16kHz and a Kinect sensor is used for RGB capture with $960 \times 540$ resolution. OptiTrack system is used to annotate the recorded sequence and provide 3D ground truth.

\section{Metrics}
\label{metrics}
There are several metrics to evaluate the performance of audio-visual speaker trackers such as OSPA \cite{schuhmacher2008consistent}, Mean Absolute Error (MAE), Track Loss Rate (TLR), MOTA \cite{bernardin2008evaluating} and MOTP \cite{bernardin2008evaluating}. \added{We provide the definition of each metric and discuss which scenario each metric can be used.}

\subsection{OSPA}

Optimal Sub-Pattern Assignment (OSPA), as defined in \cite{schuhmacher2008consistent}, operates on the reference set $ M = \left\{m_{1}, m_{2}, ..., m_{\left|M\right|}\right\}$ and the estimated set $ N = \left\{n_{1}, n_{2}, ..., n_{\left|N\right|}\right\}$. Here, $\left| \cdot \right|$ represents the length of each set.
\begin{equation}
\begin{aligned}
    &E_{\rho}^{(c)}(M, N) = \\& \left(\frac{1}{\left|N\right|}\left(\min _{\pi \in \Pi_{\left|N\right|}} \sum_{i=1}^{\left|M\right|} d^{(c)}\left(m_{i},n_{\pi(i)}\right)^{\rho}+c^{\rho}\left(\left|N\right|-\left|M\right|\right)\right)\right)^{\frac{1}{\rho}} 
\end{aligned}
\end{equation}

\noindent where $ c > 0 $ defines the largest distance and accounts for the cardinality errors, and $ \rho \geq 1 $ is the order. $\Pi_{\left|N\right|} $ denotes the set of permutations on $\left\{1, 2, ..., \left|N\right|\right\}$. \added{$\pi$ is the subset of $\Pi$}. The term $ d^{(c)}\left(m_{i},n_{\pi(i)}\right) $ is \added{defined as  $min(\Vert m_{i} - n_{\pi(i)} \Vert_{2}, c)$}.

The objective of this metric is to determine the optimal assignment of points within sets $ M $ and $ N $, effectively associating them while calculating the $\rho$-order distance between the matched points. Points within set $ N $ that remain unassociated contribute to cardinality errors.

\added{As OSPA is calculated using the best matching pairs between $ M $ and $ N $, it is needed when determining the ID information is not important.}

\subsection{MAE}
MAE is defined as follows:
\begin{equation}
    \varepsilon=\frac{1}{|K| T} \sum_{i=1}^{|K|} \sum_{t=1}^{T}\left\|\hat{\mathbf{x}}_{t, i}-\mathbf{x}_{t, i}\right\|_{2}
\end{equation}

\noindent where $|K|$ is the number of targets, $T$ is the number of time frames, $\hat{\mathbf{x}}_{t, i}$ is the predicted position and $\mathbf{x}_{t, i}$ is the ground truth position. \added{The unit of MAE is meters in 3D and pixel distance in 2D.}

\added{MAE is the commonly-used metric for evaluating the distance error and can be applied in most scenario.}

\subsection{TLR}
TLR is defined as the percentage \added{(\%)} of unsuccessful tracking over all frames. For 2D tracking on the image plane, the unsuccessful tracking is defined as that MAE is beyond $1/\lambda_{2D}$ of the length of the image diagonal. For 3D tracking, the unsuccessful tracking is defined as the case where the MAE is beyond $\lambda_{3D}$ centimeters. Typically, $\lambda_{2D}$ is often set as \num{15} and $\lambda_{3D}$ is often set as \num{30} cm. 

\added{TLR can be used in surveillance system to check whether the target is captured.}

\subsection{MOTA}
MOTA is the multiple object tracking accuracy, which is defined as follows:
\begin{equation}
    \mathrm{MOTA}=100 \% \times\left(1-\frac{\sum_{t}\left(\mathrm{FN}_{t}+\mathrm{FP}_{t}+\mathrm{MM}_{t}\right)}{\sum_{t} G _{t}}\right)
\end{equation}

\noindent where $FN$ is the number of false negative targets, or the missing targets, $FP$ is the number of false positive targets, $MM$ is the number of mismatches, and $G$ is the number of ground truth targets.

\subsection{MOTP}

MOTP stands for the multiple object tracking precision and is defined as the average localization errors over matched targets:
\begin{equation}
    \mathrm{MOTP}=\frac{\sum_{i, t} e_{t}^{i}}{\sum_{t} m_{t}}
\end{equation}

\noindent where $e_{t}^{i}$ is the Euclidean distance between the $i$-th predicted target and the matched ground truth, and $m_{t}$ is the number of matched pairs at time step $t$.

\added{Both MOTA and MOTP can be used in the scenario where the number of targets are large and ID switching is frequent.}

In addition to the aforementioned metrics, other metrics such as higher order tracking accuracy (HOTA) \cite{luiten2021hota}, most tracked targets (MT) and most lost targets (ML) are also used. The details can be found in the MOT Challenge\footnote{https://motchallenge.net/results/MOT15/}.

\subsection{Metrics for Measuring Computational Efficiency}

The metrics mentioned above are performance metrics. The computation time of trackers is also important as some applications have real-time requirements \cite{milan2017online}. For evaluating the tracker's processing velocity, Frame per Second (FPS) is a good choice. Floating Point Operations Per Second (FLOPS) can also be used.

\section{Performance Comparison of Different Trackers}
\label{methods}
\subsection{Bayesian filters}
\begin{table*}[htbp]
\caption{Overview of methods (\textbf{No. Spk} denotes the maximum number of speakers. \textbf{H} denotes the color histogram, \textbf{FD} denotes the number of face detection, \textbf{DLM} denotes dictionary learning measurement and \textbf{SR} denotes Self-Recorded. In the column of \textbf{Class}, 1 denotes Bayesian Filter with Parametric-based Measurements, 2 denotes Variational Bayesian Inference, 3 denotes Differentiable Bayesian Filter, 4 denotes Bayesian Filter with Deep Learning-Based Measurements, 5 denotes Deep Learning-Based Tracker and 6 denotes Traditional Learning-Based Tracker.)}
\centering
\small
\setlength{\tabcolsep}{0.8mm}{
\begin{tabular}{|c|c|c|c|c|c|c|c|c|c|c|}
\hline
\textbf{Ref} & \textbf{Publication} & \textbf{Tracker Name Abbr.} & \textbf{Year} & \textbf{Class} & \textbf{Backbone} & \textbf{Audio Feature} & \textbf{Visual Feature} & \textbf{Output} & \multicolumn{1}{c|}{\textbf{Dataset}} & \textbf{No. Spk} \\ \hline
\cite{d2012person}         & IET      & -            & 2012   &  1     & EKF                & GCC-PHAT                    & Mean Shift               & 2D                     & SR                                & 2              \\ \hline
\cite{barnard2014robust}         & TMM   & -               & 2014   & 6       & PF                & DOA                    & DLM               & 3D                     & \begin{tabular}[c]{@{}l@{}}AV16.3\\ CLEAR\\ EPFL\end{tabular}                                & 5              \\ \hline
\cite{kilicc2014audio}         & TMM   & AV-A-PF               & 2015   & 1       & PF                & DOA                    & H               & 2D                     & AV16.3                                & 3              \\ \hline
\cite{kilicc2016mean}         & TMM  & AVMS
SMC-PHD               & 2016    & 1       & SMC-PHD           & DOA                    & H               & 2D                     & \begin{tabular}[c]{@{}l@{}}AV16.3\\ AMI\\ CLEAR\end{tabular}                      & 4              \\ \hline
\cite{qian20173d}         & ICASSP   &   AV3D            & 2017  & 4        & PF           & GCF                    & FD, H               & 3D                     & AV16.3                    & 1              \\ \hline
\cite{qian20183d}         & ICASSP   & -               & 2018     & 4       & PF           & GCF                    & FD, H               & 3D                     & \begin{tabular}[c]{@{}l@{}}AV16.3\\ CAV3D \end{tabular}                    & 3              \\ \hline
\cite{qian2019multi}         & TMM   &  AV3T              & 2019     & 4       & PF           & GCF                    & FD, H               & 2D, 3D                     & \begin{tabular}[c]{@{}l@{}}AV16.3\\ CAV3D \end{tabular}                     & 3             \\\hline
\cite{liu2019audio}         & TMM   &    AVPF SMC-PHD           & 2019     & 4       & SMC-PHD           & DOA                    & FD, H               & 2D, 3D                     & \begin{tabular}[c]{@{}l@{}}AV16.3\\ AVDIAR \\CLEAR \end{tabular}                     & 4             \\\hline
\cite{liu20193d}         & ICIP    & 2LPF              & 2019     &4     & PF           & SSM                    & H               & 3D                     & AV16.3                    & 3             \\\hline
\cite{ban2019variational}         & TPAMI  &     VAVIT           & 2019    &2      & EM           & DP-RTF                    & FD               & 2D                     & \begin{tabular}[c]{@{}l@{}}AV16.3\\ AVDIAR \end{tabular}                     & 4             \\\hline
\cite{lin2020audio}         & \begin{tabular}[c]{@{}l@{}}INTER-\\ SPEECH \end{tabular}      & AV-GLMB            & 2020   &4       & GLMB           & GCF                    & FD               & 2D, 3D                     & AV16.3                    & 3             \\\hline
\cite{schymura2020dynamic}         & ICASSP    & DKF              & 2020    &3      & KF           & SRP-PHAT                    & Facial Landmarks               & DOA                     & SR                    & 1             \\\hline
\cite{liu20213d}         & ICPR              & -    & 2021   &4       & PF           & GCF                    & FD, H               & 3D                     & \begin{tabular}[c]{@{}l@{}}AV16.3\\ CAV3D \end{tabular}                     & 3             \\\hline
\cite{li2022multi}         & AAAI      & MPT            & 2022     &4     & PF           & stGCF                    & Siamese Network               & 2D                     & AV16.3                    & 1             \\\hline
\cite{zhao2022audio}         & ICASSP        & AV-PMBM          & 2022   &4       & PMBM           & DOA                    & FD               & 2D                     & AV16.3                    & 3             \\\hline
\cite{zhao2022audio2}         & \begin{tabular}[c]{@{}l@{}}INTER-\\ SPEECH \end{tabular}     & -             & 2022     &4      & PMBM          & GCF                    & FD               & 3D                     & AV16.3                    & 3             \\\hline
\cite{qian2022audio}         & TASLP    & CMAF              & 2023     &5     & Transformer           & GCC-PHAT                    & FD               & DOA                     & AVRI                    & 1             \\\hline
\cite{sanabria2023audiovisual}         & SENSORS     & -             & 2023      &4     & PF           & GCC-PHAT                    & FD, H               & 2D, 3D                     & \begin{tabular}[c]{@{}l@{}}AV16.3\\ CAV3D \end{tabular}                     & 3             \\\hline
\cite{liu2023labelled}         & TMM           & LPF     & 2023    &4       & SMC-PHD          & DOA                    & FD               & 2D                     & \begin{tabular}[c]{@{}l@{}}AV16.3\\ AVDIAR \\CLEAR \end{tabular}                    & 4             \\\hline
\end{tabular}}
\label{overview}
\end{table*}

We summarize the MAE results of current trackers on AV16.3 dataset. \added{We classifies the trackers according to the principles in Table \ref{overview}.} The results on multiple speakers sequences are shown in Table \ref{multiple}. In \cite{kilicc2014audio}, the number of speakers is known as a priori. This algorithm used color histograms as visual measurements and focused on the particles near the DOA lines. The PF filter in \cite{kilicc2014audio} achieved the best performance in most \added{multiple-speaker sequences in AV16.3}. As a following work, \cite{kilicc2016mean} solved the problem of audio-visual speaker tracking with an SMC-PHD filter, which does not need to know the number of speakers as a priori. Mean-shift algorithm \cite{comaniciu2003kernel} was further employed to move the particles closer to the speakers' locations. In \cite{qian2021audio}, GCF is used to calculate sound source positions as audio measurements and utilized face detectors to derive the month positions as visual measurements. GLMB filter was employed to fuse the two modalities and generate the trajectories. In \cite{barnard2014robust}, dictionary learning is used to model the appearance of speakers and used PF as the tracking framework. DOA lines are used in this algorithm for the initialization of particle positions. The MAE results on single speaker sequence are shown in Table \ref{single}. In \cite{qian20173d}, an adaptive PF is designed where the covariance of the measurement likelihood function can change dynamically according to the reliability of the measurements. In \cite{liu20193d}, a two-layer PF is designed for 3D single-speaker tracking. The designed two-layer architecture increases the particle diversity. The particle weights are adjusted according to the confidence of audio and visual modalities. In \cite{liu20213d}, a novel PF is proposed, which performs audio azimuth relocation and audio-visual azimuth-elevation relocation. Face detection is employed to estimate the distance. The measurement likelihood is derived based on the angle likelihood and the distance likelihood. \added{Multi-modal perception tracker (MPT) \cite{li2022multi} is the first attempt to use deep learning techniques for tracking, which uses cross-modal self-supervised learning to determine the importance of different modalities. }

We also summarize the performance of trackers on CAV3D dataset \cite{qian2019multi}. The metrics on the image plane and 3D space can be found in TABLE \ref{cav3d_2d}
 and TABLE \ref{cav3d_3d}, respectively. We report the performance on single object tracking (SOT) sequences and multiple object tracking (MOT) sequences separately. SOT sequences refer to sequence 06 $\sim$ 13 and sequence 20. MOT sequences refer to sequence 22 $\sim$ 26. It is found that the MAE on CAV3D dataset is higher than that on AV16.3 dataset. As CAV3D is more challenging, it has stronger reverberation and more scenarios of occlusions and out-of-views.
\begin{table}[htbp]
\centering
\caption{Tracking results on multiple speaker sequences of AV16.3 datasets. The bold numbers indicate the best tracker for a given sequence and a given camera. The category of trackers (class) is classified using the principles in Table \ref{overview}. }
\setlength{\tabcolsep}{1mm}{
\begin{tabular}{|l|c|cccc|}
\hline
\multicolumn{1}{|c|}{\multirow{2}{*}{\textbf{Sequence}}} & \multirow{2}{*}{\textbf{Camera}} & \multicolumn{4}{c|}{\textbf{Trackers (Class)}}                                                                                                                                                                     \\ \cline{3-6} 
\multicolumn{1}{|c|}{}                                   &                                  &  \multicolumn{1}{c|}{\begin{tabular}[c]{@{}c@{}}AV-A-PF \\ (1)\cite{kilicc2014audio}\end{tabular}} & \multicolumn{1}{c|}{\begin{tabular}[c]{@{}c@{}}AVMS\\  SMC-PHD \\ (1)\cite{kilicc2016mean}\end{tabular}} & \multicolumn{1}{c|}{\begin{tabular}[c]{@{}c@{}}AV-GLMB \\ (4)\cite{lin2020audio}\end{tabular}} & \begin{tabular}[c]{@{}c@{}}Tracker\\ (1)\cite{barnard2014robust}\end{tabular} \\ \hline
\multicolumn{1}{|c|}{\multirow{3}{*}{Seq18-2p-0101}}     & Cam1                             & \textbf{14.31}               & -                                                                            & 15.7                         & -                                                             \\ \cline{2-2}
\multicolumn{1}{|c|}{}                                   & Cam2                             & 11.66                        & -                                                                            & \textbf{10.9}                & -                                                             \\ \cline{2-2}
\multicolumn{1}{|c|}{}                                   & Cam3                             & 15.80                        & -                                                                            & \textbf{6.3}                 & -                                                             \\ \hline
\multirow{3}{*}{Seq19-2p-0101}                           & Cam1                             & \textbf{11.88}               & -                                                                            & 15.3                         & -                                                             \\ \cline{2-2}
                                                         & Cam2                             & \textbf{9.62}                & -                                                                            & 11.6                         & -                                                             \\ \cline{2-2}
                                                         & Cam3                             & 12.08                        & -                                                                            & \textbf{5.4}                 & -                                                             \\ \hline
\multirow{3}{*}{Seq24-2p-0111}                           & Cam1                             & \textbf{9.95}                & 13.93                                                                        & 16.5                         & 22.28                                                         \\ \cline{2-2}
                                                         & Cam2                             & \textbf{8.85}                & 14.97                                                                        & 10.6                         & 17.60                                                         \\ \cline{2-2}
                                                         & Cam3                             & 10.02                        & 14.12                                                                        & \textbf{7.0}                 & 28.18                                                         \\ \hline
\multirow{3}{*}{Seq25-2p-0111}                           & Cam1                             & \textbf{14.78}               & 15.72                                                                        & 17.7                         & 21.49                                                         \\ \cline{2-2}
                                                         & Cam2                             & \textbf{7.70}                & 13.93                                                                        & 10.8                         & 19.17                                                         \\ \cline{2-2}
                                                         & Cam3                             & \textbf{8.93}                & 17.07                                                                        & 10.7                         & 29.35                                                         \\ \hline
\multirow{3}{*}{Seq30-2p-1101}                           & Cam1                             & \textbf{13.84}               & 16.65                                                                        & 14.8                         & 35.98                                                         \\ \cline{2-2}
                                                         & Cam2                             & \textbf{8.85}                & 14.86                                                                        & 10.4                         & 28.40                                                         \\ \cline{2-2}
                                                         & Cam3                             & \textbf{10.30}               & 19.29                                                                        & 15.7                         & 34.60                                                         \\ \hline
\multirow{3}{*}{Seq40-3p-0111}                           & Cam1                             & \textbf{12.38}               & -                                                                            & -                            & -                                                             \\ \cline{2-2}
                                                         & Cam2                             & \textbf{12.04}               & -                                                                            & -                            & -                                                             \\ \cline{2-2}
                                                         & Cam3                             & \textbf{11.30}               & -                                                                            & -                            & -                                                             \\ \hline
\multirow{3}{*}{Seq45-3p-1111}                           & Cam1                             & \textbf{16.35}               & 22.95                                                                        & -                            & -                                                             \\ \cline{2-2}
                                                         & Cam2                             & \textbf{17.22}               & 21.47                                                                        & -                            & -                                                             \\ \cline{2-2}
                                                         & Cam3                             & \textbf{13.84}               & 22.43                                                                        & -                            & -                                                             \\ \hline
\end{tabular}}
\label{multiple}
\end{table}

\begin{table}[htbp]
\centering
\scriptsize
\caption{Tracking results on single speaker sequences of AV16.3 datasets. The bold numbers indicate the best tracker for a given sequence and a given camera. The category of trackers (class) is classified using the principles in Table \ref{overview}.}
\setlength{\tabcolsep}{1mm}{
\begin{tabular}{|lc|ccccc|}
\hline
\multicolumn{1}{|c|}{\multirow{2}{*}{\textbf{Sequence}}}      & \multirow{2}{*}{\textbf{Camera}} & \multicolumn{5}{c|}{\textbf{Trackers (Class)}}                                                                                                                                                                                                                                                                                                                                          \\ \cline{3-7} 
\multicolumn{1}{|c|}{}                                        &                                  & \multicolumn{1}{c|}{\begin{tabular}[c]{@{}c@{}}AV-A-PF\\ (1)\cite{kilicc2014audio}\end{tabular}} & \multicolumn{1}{c|}{\begin{tabular}[c]{@{}c@{}}AV3D \\ (4)\cite{qian20173d}\end{tabular}} & \multicolumn{1}{c|}{\begin{tabular}[c]{@{}c@{}}2LPF \\ (4)\cite{liu20193d}\end{tabular}} & \multicolumn{1}{c|}{\begin{tabular}[c]{@{}c@{}}MPT \\ (4)\cite{li2022multi}\end{tabular}} & \begin{tabular}[c]{@{}c@{}}Tracker\\ (4)  \cite{liu20213d}\end{tabular} \\ \hline
\multicolumn{1}{|c|}{\multirow{3}{*}{Seq08-2p-0100}} & Cam1                             & 10.75                                                                       & 4.31                                                                     & 3.32                                                                     & 3.67                                                                    & \textbf{3.01}                                                     \\ \cline{2-2}
\multicolumn{1}{|c|}{}                                        & Cam2                             & 7.33                                                                        & 4.66                                                                     & 3.08                                                                     & 3.58                                                                    & \textbf{2.30}                                                     \\ \cline{2-2}
\multicolumn{1}{|c|}{}                                        & Cam3                             & 9.85                                                                        & 5.34                                                                     & 3.47                                                                     & \textbf{3.43}                                                           & 3.59                                                              \\ \hline
\multicolumn{1}{|l|}{\multirow{3}{*}{Seq11-2p-0100}} & Cam1                             & 14.66                                                                       & 8.15                                                                     & 6.15                                                                     & 6.77                                                                    & \textbf{5.43}                                                     \\ \cline{2-2}
\multicolumn{1}{|l|}{}                                        & Cam2                             & 14.01                                                                       & 7.48                                                                     & 5.58                                                                     & \textbf{4.55}                                                           & 4.60                                                              \\ \cline{2-2}
\multicolumn{1}{|l|}{}                                        & Cam3                             & 13.96                                                                       & 6.64                                                                     & 3.86                                                                     & \textbf{3.84}                                                           & 6.28                                                              \\ \hline
\multicolumn{1}{|l|}{\multirow{3}{*}{Seq12-2p-0100}} & Cam1                             & 12.49                                                                       & 6.86                                                                     & \textbf{4.11}                                                            & 4.67                                                                    & 4.23                                                              \\ \cline{2-2}
\multicolumn{1}{|l|}{}                                        & Cam2                             & 10.81                                                                       & 10.67                                                                    & 5.39                                                                     & \textbf{4.84}                                                           & 4.53                                                              \\ \cline{2-2}
\multicolumn{1}{|l|}{}                                        & Cam3                             & 11.86                                                                       & 9.71                                                                     & 5.65                                                                     & \textbf{3.78}                                                           & 4.25                                                              \\ \hline
\multicolumn{2}{|c|}{\textbf{Average}}                                                           & 11.74                                                                       & 7.09                                                                     & 4.51                                                                     & 4.34                                                                    & \textbf{4.25}                                                     \\ \hline
\end{tabular}}
\label{single}
\end{table}

\begin{table}[htbp]
\centering
\caption{2D metrics on CAV3D datasets. The bold numbers indicate the best tracker. The category of trackers (class) is classified using the principles in Table \ref{overview}.}
\setlength{\tabcolsep}{1mm}{
\begin{tabular}{|c|c|ccc|}
\hline
\multicolumn{1}{|l|}{\multirow{2}{*}{\textbf{Sequences}}} & \multicolumn{1}{l|}{\multirow{2}{*}{\textbf{Metrics}}} & \multicolumn{3}{c|}{\textbf{Trackers (Class)}}                                          \\ \cline{3-5} 
\multicolumn{1}{|l|}{}                           & \multicolumn{1}{l|}{}                         & \multicolumn{1}{c|}{Tracker (4)\cite{liu20213d}} & \multicolumn{1}{c|}{AV3T (4)\cite{qian2019multi}} & GAVT (4)\cite{sanabria2023audiovisual}           \\ \hline
\multirow{2}{*}{SOT}                             & TLR                                           & \textbf{2.50}             & 7.00                      & 13.93          \\ \cline{2-2}
                                                 & MAE                                           & \textbf{12.00}            & 16.50                     & 26.76          \\ \hline
\multirow{2}{*}{MOT}                             & TLR                                           & -                         & \textbf{11.20}            & 21.01          \\ \cline{2-2}
                                                 & MAE                                           & -                         & 24.80                     & \textbf{13.47} \\ \hline
\end{tabular}}
\label{cav3d_2d}
\end{table}

\begin{table}[htbp]
\centering
\caption{3D metrics on CAV3D datasets. The bold numbers indicate the best tracker. The category of trackers (class) is classified using the principles in Table \ref{overview}.}
\setlength{\tabcolsep}{1mm}{
\begin{tabular}{|c|c|ccc|}
\hline
\multicolumn{1}{|l|}{\multirow{2}{*}{\textbf{Sequences}}} & \multicolumn{1}{l|}{\multirow{2}{*}{\textbf{Metrics}}} & \multicolumn{3}{c|}{\textbf{Trackers}}                                          \\ \cline{3-5} 
\multicolumn{1}{|l|}{}                           & \multicolumn{1}{l|}{}                         & \multicolumn{1}{c|}{Tracker (4)\cite{liu20213d}} & \multicolumn{1}{c|}{AV3T (4)\cite{qian2019multi}} & GAVT (4)\cite{sanabria2023audiovisual}           \\ \hline
\multirow{2}{*}{SOT}                             & TLR                                           & \textbf{20.70}             & 31.80                      & 30.07          \\ \cline{2-2}
                                                 & MAE                                           & \textbf{0.21}            & 0.30                     & 0.29          \\ \hline
\multirow{2}{*}{MOT}                             & TLR                                           & -                         & 35.70            & \textbf{32.01}          \\ \cline{2-2}
                                                 & MAE                                           & -                         & 0.37                     & \textbf{0.32} \\ \hline
\end{tabular}}
\label{cav3d_3d}
\end{table}

\subsection{Deep Learning-Based Trackers}

In Table \ref{deep_trackers}, we summarize the performance of the tracker mentioned in Section IV.\ref{track_deep_learning} using the MOT17 private benchmark\footnote{The results are from https://motchallenge.net/results/MOT17/?det=Private}. We report MOTA and HOTA introduced in Section \ref{metrics} for accuracy evaluation. We also report ID switching for evaluating the model ability in maintaining ID consistency cross consecutive frames. It is shown C-BIoU \cite{yang2023hard} achieves the best MOTA and HOTA due to the designed buffered IOU matching strategy, which is simple but effective. MotionTrack \cite{qin2023motiontrack} has a lower ID switching number compared to the other trackers due to the short-term and long-term association module.

\subsection{Differentiable Bayesian Filers}
\added{Although different methods are evaluated on different dataset and there are no unified benchmarks for differentiable Bayesian filters, we still summarize the performance of different methods in Table }\ref{dbf_summ}. \added{The differentiable particle filter proposed in \cite{jonschkowski2018differentiable} optimize the learnable transition and update model in a end-to-end manner and the performance on KITTI \cite{geigervision} dataset outperforms the Backprop Kalman filter \cite{haarnoja2016backprop}. Following a similar idea, PFnet proposed in \cite{karkus2018particle} is applied in global localization and achieves better performance than LSTM and traditional PF. PF-LSTM and PF-GRU is proposed in \cite{ma2020particle} and is used in regression tasks such as stock index prediction and appliances energy prediction. Experimental results show that PF-LSTM and PF-GRU are superior to the vanilla LSTM and GRU.}

\begin{table}[tbp]
\centering
\caption{Performance comparisons of deep-learning-based trackers. The bold numbers indicate the best tracker.}
\setlength{\tabcolsep}{1mm}{
\begin{tabular}{|l|l|l|l|}
\hline
            & \textbf{MOTA $\uparrow$} & \textbf{HOTA$\uparrow$} & \textbf{ID Swit.$\downarrow$} \\ \hline
Trackformer \cite{meinhardt2022trackformer} & 74.1          & 57.3          & 2,829          \\ \hline
TransCenter \cite{xu2021transcenter} & 73.2          & 54.5          & 4,614          \\ \hline
TransTrack \cite{sun2020transtrack}  & 75.2          & 54.1          & 3,603         \\ \hline
ByteTrack \cite{zhang2022bytetrack}  & 80.3          & 63.1          & 2,196         \\ \hline
MotionTrack \cite{qin2023motiontrack} & 81.1        & 65.1          & \textbf{1,140}         \\ \hline
C-BIoU \cite{yang2023hard} & \textbf{82.8}          & \textbf{66.0}          & 1,194         \\ \hline
\end{tabular}}
\label{deep_trackers}
\end{table}

\begin{table*}[tbp]
\centering
\caption{Performance summarization of differentiable Bayesian filters.}
\setlength{\tabcolsep}{1mm}{
\begin{tabular}{|l|l|l|l|l|l|l|l|l|}
\hline
\textbf{Methods} & \textbf{Task}                                                                                 & \textbf{Dataset}                                        & \textbf{Metric}                                                           & \textbf{Unit}                                           & \textbf{Result}                                     & \textbf{Metric}                                             & \textbf{Unit}                                   & \textbf{Result}                                       \\ \hline
\cite{jonschkowski2018differentiable}               & visual odometry                                                                               & KITTI                                                   & translational error                                                       & meters                                                  & 0.1467                                              & rotational error                                            & degree                                          & 0.0499                                                \\ \hline
\cite{karkus2018particle}            & global localization                                                                           & House3D                                                 & RMSE                                                                      & centimeters                                                      & 40.50                                               & success rate                                                & \%                                              & 82.6\%                                                \\ \hline
\cite{ma2020particle}            & \begin{tabular}[c]{@{}l@{}}stock index prediction\\ appliances energy prediction\end{tabular} & \begin{tabular}[c]{@{}l@{}}SML2010\\ recorded dataset \cite{candanedo2017data} \end{tabular} & \begin{tabular}[c]{@{}l@{}}regression loss\\ regression loss\end{tabular} & -                                                       & \begin{tabular}[c]{@{}l@{}}1.33\\ 3.72\end{tabular} & -                                                           & -                                               & -                                                     \\ \hline
\cite{schymura2020dynamic}                 & speaker tracking                                                                              & in-house                                                & gross accuracy                                                            & \%                                                      & 50$\sim$100                                          & -                                                           & -                                               & -                                                     \\ \hline
\cite{zhao2024attention}           & speaker tracking                                                                              & \begin{tabular}[c]{@{}l@{}}in-house\\ AVRI\end{tabular} & \begin{tabular}[c]{@{}l@{}}MAE\\ MAE\end{tabular}                         & \begin{tabular}[c]{@{}l@{}}degree\\ degree\end{tabular} & \begin{tabular}[c]{@{}l@{}}4.40\\ 8.70\end{tabular} & \begin{tabular}[c]{@{}l@{}}accuracy\\ accuracy\end{tabular} & \begin{tabular}[c]{@{}l@{}}\%\\ \%\end{tabular} & \begin{tabular}[c]{@{}l@{}}95.17\\ 78.05\end{tabular} \\ \hline
\end{tabular}}
\label{dbf_summ}
\end{table*}

\section{Future Directions}
\label{future}
\subsection{Audio-Visual Multiple Speaker Tracking with Speech Separation}
Audio-visual tracking can assist speech separation. Compared to the deep learning based methods such as \cite{luo2018tasnet} and \cite{wang2018voicefilter}, the detection, tracking and filtering (DTF) framework for speech separation can be adapted to the varying numbers of speakers without training on large-scale datasets. In \cite{ong2022audio}, GLMB is employed to generate speaker trajectories with audio and visual measurements. The generalized side-lobe canceller (GSC) \cite{nordholm2014broadband} is implemented based on the trajectories to perform online speech separation. In \cite{subramanian2021directional}, an end-to-end far-field speech recognition system is proposed integrating localization, separation and ASR. The localization part gives the interpretation and improves the performance.

Speech separation can help the obtaining of the audio measurements. In \cite{wang2018robust}, deep neural networks are used to calculate time-frequency masking, aiming to obtain the clean phase for DOA estimation. In \cite{subramanian2022deep}, the speech separation method is used for DOA estimation. And the obtained DOA improves the performance of downstream tasks such as ASR. In \cite{chen2023locselect}, voicefilter \cite{wang2018voicefilter} is used to separate the target voice before localizing the target speaker.


\subsection{Audio-Visual Multiple Speaker Tracking in Distributed Scenarios}
\label{distributed}
Audio-visual tracking utilizes more modalities to improve tracking accuracy than past tracking systems, which use a single modality. However, most of the existing works for audio-visual tracking only utilize one microphone array and one camera, and those sensors are often regarded as one node. Therefore, when a sensor in the node cannot work as well as expected, the unreliable measurements from that sensor will degrade the tracking accuracy. The simple idea for solving this problem is to increase the number of nodes to obtain global estimates, which can mitigate the impacts of errors in each node's estimates. Thus,  there has been a growing interest in the development of tracking systems using distributed sensors, which have the potential to enhance tracking accuracy and reliability \cite{bhatti2009survey}. Recently, several distributed filters have been proposed for tracking in the distributed scenario \cite{8902553, rezaei2020event, ma2019unscented, wu2022partial}.

\subsection{Audio-Visual Multiple Speaker Tracking in Egocentric Scenarios}
Recently tasks in egocentric scenarios arise increasing interest as the egocentric scenario mimics the similar way as humans explore the world. There are some large-scale datasets such as Epic-kitchen \cite{damen2018scaling} and Ego4D \cite{ grauman2022ego4d} for egocentric perception. In these datasets, one person wears egocentric equipment (the wearer) such as cameras and microphones to record their daily life. Audio-visual speaker tracking in egocentric scenarios is beneficial for audio-visual navigation and human robot interaction. However, there are some challenges, which differ from the conventional audio-visual speaker tracking scenarios, including motion blur, speaker disappearance due to the movement of the wearer and occlusions. For this task, the Easycom dataset \cite{donley2021easycom} is proposed for audio-visual active speaker localization and tracking in the egocentric scenario. In \cite{zhao2023audio}, a simulated is proposed, in which the speaker moves more frequently and more out-of-view scenarios are included compared to the Easycom dataset.

\subsection{Prompt-Based Target Speaker Tracking}
Text and audio signals can be used as prompts to describe the target sound event. Text prompt is used in \cite{wang2021towards} for target object tracking. In \cite{jiang2023prompt}, target speech diarization is proposed and text can be used to indicate the target speaker such as the female speaker or the dominant speaker. Audio is used in \cite{chen2023locselect} \added{as a condition} to provide reference speech for target speaker localization. For the task of audio-visual speaker tracking, audio or text prompts can be used to describe the target speaker as well. The prompt-based tracking task provides a more user-friendly way of human-computer interaction and can be used in monitoring systems.

\section{Conclusions}

In this paper, we conduct a comprehensive literature review on audio-visual speaker tracking. We recall the existing methods for obtaining audio and visual measurements. And we introduce the Bayesian filters. We discuss the new techniques, especially the deep learning based methods such as the learning-based features and differentiable Bayesian filters. Though the development of audio-visual speaker tracking has been progressive during the past few years, there remain some problems. Firstly, most current trackers do not evaluate the computational complexities. However, some downstream tasks such as speech enhancement and monitoring have the requirements of real-time tracking. In addition, there still a lack of large-scale datasets for audio-visual multiple speakers tracking, for which deep learning techniques are not fully explored in this traditional task.

\section{Author Contributions}
\textbf{Jinzheng Zhao}: Writing – original draft. \textbf{Yong Xu}: Project administration; Review and editing. \textbf{Xinyuan Qian}: Review and editing. \textbf{Davide Berghi}: Writing - Part of contents in Section \ref{intro} and Section II.B.\ref{teacherstudent}. \textbf{Peipei Wu}: Writing - Section VIII.\ref{distributed}. \textbf{Meng Cui}: Discussion. \textbf{Jianyuan Sun}: Discussion. \textbf{Philip J.B. Jackson}: Discussion and analysis. \textbf{Wenwu Wang}: Project administration; Review and editing. Funding acquisition.

\bibliographystyle{ieeetr}
\bibliography{sample-base}

\end{document}